\documentclass[prd,twocolumn,preprintnumbers,superscriptaddress,nofootinbib]{revtex4}
\pdfoutput=1
\usepackage[a4paper, hdivide={1.91cm,,1.165cm}, vdivide={1.83cm,,3.2cm}]{geometry}

\usepackage{amsmath}
\usepackage{bbold}
\usepackage{graphicx}
\usepackage{xspace}
\usepackage{color}
\usepackage{units}
\usepackage{slashed}
\usepackage[hyperfootnotes=false]{hyperref}
\usepackage{gensymb}

\def \L {\mathcal{L}} 

\def \epsilon {\varepsilon} 

\newcommand{\matrixx}[1]{\begin{pmatrix} #1 \end{pmatrix}} 
\newcommand{\M}{\mathcal{M}}
\newcommand{\hc}{\ensuremath{\text{h.c.}}}
\newcommand{\Br}{\text{Br}}
\newcommand{\tr}{\text{tr}}
\newcommand{\diag}{\text{diag}}
\newcommand{\dd}{\mathrm{d}}

\def \Re {\text{Re}} 

\def \Im {\text{Im}} 

\def \i {\mathrm{i}\mkern1mu} 

\allowdisplaybreaks

\begin{document}

\title{ Neutrino Lines from Majoron Dark Matter
}

\preprint{ULB-TH/17-01}

\author{Camilo Garcia-Cely}
\email{Camilo.Alfredo.Garcia.Cely@ulb.ac.be}
\affiliation{Service de Physique Th\'eorique, Universit\'e Libre de Bruxelles, Boulevard du Triomphe, CP225, 1050 Brussels, Belgium}

\author{Julian \surname{Heeck}}
\email{Julian.Heeck@ulb.ac.be}
\affiliation{Service de Physique Th\'eorique, Universit\'e Libre de Bruxelles, Boulevard du Triomphe, CP225, 1050 Brussels, Belgium}

\hypersetup{
    pdftitle={Neutrino Lines from Majoron Dark Matter},
    pdfauthor={Julian Heeck, Camilo Garcia-Cely}
}


\begin{abstract}

Models with spontaneously broken global lepton number can lead to a pseudo-Goldstone boson as a long-lived dark matter candidate. Here we revisit the case of singlet majoron dark matter and discuss multiple constraints. For masses above MeV, this model could lead to a detectable flux of monochromatic mass-eigenstate neutrinos, which have flavor ratios that depend strongly on the neutrino mass hierarchy. We provide a convenient parametrization for the loop-induced majoron couplings to charged fermions that allows us to discuss three-generation effects such as lepton flavor violation. These couplings are independent of the low-energy neutrino parameters but can be constrained by the majoron decays into charged fermions.

\end{abstract}

\maketitle


\section{Introduction}

The observation of neutrino oscillations has raised the question why neutrino masses are so much smaller than all other known masses. The most-studied solution to this puzzle comes in the form of the seesaw mechanism~\cite{Minkowski:1977sc}, where heavy right-handed neutrinos suppress active-neutrino masses with respect to the electroweak scale. As a bonus, these heavy neutrinos can dynamically generate a baryon asymmetry in the early Universe via leptogenesis~\cite{Fukugita:1986hr}, thus solving a further problem of the Standard Model (SM). An inherent feature of the seesaw mechanism is the self-conjugate Majorana nature of neutrinos, which implies that the anomaly-free global $U(1)_{B-L}$ symmetry of the SM has to be broken by two units. Following the success of spontaneous symmetry breaking in particle physics, one can easily imagine that also this $B-L$ symmetry is broken \emph{spontaneously}, resulting in a Nambu--Goldstone boson named majoron~\cite{Chikashige:1980ui,Schechter:1981cv}. Gravitational or explicit breaking terms then typically generate a mass term, making the majoron a pseudo-Goldstone boson. Since the majoron has couplings that are suppressed by the $B-L$ breaking scale, i.e.~the seesaw scale, it can easily be long-lived enough to form the dark matter (DM) of our Universe~\cite{Rothstein:1992rh,Berezinsky:1993fm,Lattanzi:2007ux,Bazzocchi:2008fh,Frigerio:2011in,Lattanzi:2013uza,Queiroz:2014yna, Wang:2016vfj}. The CP-even partner of the majoron can on the other hand be used to drive inflation~\cite{Boucenna:2014uma,Higaki:2014dwa}, although this is not the focus here.
Since the seesaw scale is far above the electroweak scale for DM stability reasons, leptogenesis will be hardly modified by the majoron~\cite{Sierra:2014sta}.

The most salient and well-studied indirect-detection signature of majoron DM is its decay into two photons, which most prominently arises in cases where one identifies the majoron with the axion~\cite{Mohapatra:1982tc,Langacker:1986rj,Ballesteros:2016euj,Ballesteros:2016xej}, for which new particles have to be introduced to create a color anomaly of the $U(1)$, often accompanied by an electromagnetic anomaly as well~\cite{DiLuzio:2016sbl}. 
In this article we focus on other possible signatures, most notably from the tree-level decays into neutrinos and from the one-loop decays into charged fermions~\cite{Lattanzi:2007ux,Bazzocchi:2008fh,Frigerio:2011in}. To this effect we provide a simple parametrization of the majoron couplings to the three generations of fermions that also allows us to discuss lepton flavor violation (LFV) and perturbativity.

Owing to its tree-level coupling, the key feature of majoron DM is arguably its two-body decay into monochromatic neutrinos, a topic that has received a lot of attention in recent years in its own right~\cite{Lindner:2010rr,Gustafsson:2013gca,Aisati:2015ova,Arina:2015zoa,Queiroz:2016zwd}.
In fact, since neutrinos are the least-detectable SM particles, any limit on their flux automatically provides a model-independent lower bound on the DM lifetime~\cite{PalomaresRuiz:2007ry}. (The same arguments apply to DM annihilations~\cite{Beacom:2006tt,Yuksel:2007ac,PalomaresRuiz:2007eu}.) 
Majorons are a well-motivated DM candidate that can lead to observable monochromatic neutrino fluxes for energies between MeV and 10\,TeV.
For energies below the electroweak scale, these neutrino lines do not receive Bremsstrahlung corrections that could otherwise lead to observable gamma-ray fluxes~\cite{Kachelriess:2007aj,Bell:2008ey,Queiroz:2016zwd}, so neutrino detectors have unique detection possibilities. Experiments that are sensitive to MeV-scale supernova neutrinos, most prominently Borexino~\cite{Bellini:2010gn}, KamLAND~\cite{Collaboration:2011jza}, and Super-Kamiokande (SK)~\cite{Gando:2002ub,Zhang:2013tua}, can thus be used as DM detectors as well.

This article is organized as follows: in Sec.~\ref{sec:model} we provide an introduction to the singlet majoron model, its couplings and decay modes. In particular, we introduce a compact parametrization for the one-loop induced majoron couplings to charged fermions that is invaluable to study majorons. In Sec.~\ref{sec:DM} we discuss the signatures of majoron DM, split into neutrino signatures (Sec.~\ref{sec:neutrinos}) and visible decay modes (Sec.~\ref{sec:visible}). Low-energy constraints from e.g.~LFV that are also relevant if the majoron is not DM are presented in Sec.~\ref{sec:lfv}. Finally, we conclude in Sec.~\ref{sec:conclusion}.
Appendix~\ref{sec:apA} is devoted to a discussion of neutrino flavor ratios after propagation, highlighting the differences between neutrino production from electroweak and majoron interactions.

\section{Singlet Majoron Model}
\label{sec:model}

We know from neutrino-oscillation experiments that at least two neutrinos are massive, with sub-eV mass splitting. If neutrinos are Majorana particles, this implies that lepton number $U(1)_L$ (or $U(1)_{B-L}$) is broken by two units, and if this breaking is spontaneous we expect a Goldstone boson, the majoron $J$~\cite{Chikashige:1980ui,Schechter:1981cv}. We will restrict ourselves to the \emph{singlet} majoron model, where an SM-singlet complex scalar $\sigma = (f+\sigma^0 + \i J)/\sqrt{2}$ with $L(\sigma) = -2$ couples to three right-handed neutrinos $N_R$,
\begin{align}
\L = -\overline{L} y N_R H -\tfrac12\overline{N}_R^c\lambda  N_R \sigma +\hc ,
\label{eq:lagrangian}
\end{align}
with the lepton (scalar) doublet $L$ ($H$) and the Yukawa matrices $y$ and $\lambda$. A generalization to arbitrarily many right-handed neutrinos is straightforward and will not change the discussion~\cite{Heeck:2012fw}.
Spontaneous symmetry breaking at the scale $f$ gives rise to the right-handed Majorana mass matrix $\mathcal{M}_R =  f \lambda/\sqrt2$, diagonal without loss of generality. Electroweak symmetry breaking, $\langle H\rangle = (v/\sqrt2,0)^T$, introduces a mixing between the left and right-handed neutrinos via the Dirac mass matrix $m_D = y v/\sqrt2$. The full Majorana mass matrix in the basis $(\nu_L^c, N_R) = V n_R$ is then
\begin{align}
\matrixx{0 & m_D \\ m_D^T & \M_R} = V^* \diag (m_1,\dots, m_6) V^\dagger\,,
\label{eq:neutrino_mass_matrix}
\end{align}
where $V$ is the $6\times 6$ mixing matrix to the states $n_R$, which form the Majorana mass eigenstates $n = n_R + n_R^c$. The relevant couplings of $J$, $Z$, and $W^-$ can be rewritten in terms of these mass eigenstates as~\cite{Pilaftsis:1993af}
\begin{align}
\begin{split}
\L_J &= -\frac{\i J }{2f}\sum_{i,j=1}^6 \overline{n}_i \left[ \gamma_5 (m_i + m_j)(\tfrac12 \delta_{ij} - \Re C_{ij}) \right. \\
&\left.\hspace{3.3cm}+ \i (m_i - m_j) \Im C_{ij}\right] n_j \,, \label{eq:Jnunu}
\end{split}\\
\L_Z &= -\frac{g_w}{4 \cos\theta_w} \sum_{i,j=1}^6\overline{n}_i \slashed{Z}\left[\i\Im C_{ij} - \gamma_5 \Re C_{ij}\right] n_j\,,\\
\L_W &= -\frac{g_w}{2\sqrt2} \sum_{i,j=1}^6\overline{\ell}_i B_{\ell_i j}\slashed{W}^- (1 - \gamma_5) n_j + \hc\,,
\end{align}
where 
\begin{align}
C_{ij} \equiv \sum_{k=1}^3 V_{ki} V^*_{kj}\,, &&
B_{\ell_i j} \equiv \sum_{k=1}^3 U^\ell_{\ell_i k} V^*_{kj}\,.
\end{align}
Here, $U^\ell$ is a unitary mixing matrix from the diagonalization of the charged-lepton mass matrix which we can assume to be the identity matrix without loss of generality. The neutrino couplings to the CP-even scalars can be found in Ref.~\cite{Pilaftsis:1993af} but are of no importance here. We will assume $\sigma^0$ to be very heavy, $m_{\sigma^0}\sim f \gg v$, and essentially decoupled from the SM to simplify the discussion. It could however be used as an inflaton~\cite{Boucenna:2014uma,Higaki:2014dwa}, which has little impact on the discussion in this article.

We will further assume the majoron to be massive, i.e.~a \emph{pseudo}-Goldstone boson. The mass could arise because of (quantum-)gravity effects~\cite{Akhmedov:1992hi,Rothstein:1992rh}, heavy chiral fields that render $B-L$ anomalous (which could make $J$ an axion and identify the seesaw scale with the Peccei--Quinn scale~\cite{Mohapatra:1982tc,Langacker:1986rj,Ballesteros:2016euj,Ballesteros:2016xej}) or simply because of explicit breaking in the Lagrangian~\cite{Gu:2010ys,Frigerio:2011in}. The actual mechanism is not important for our analysis, its main impact will be on the production mechanism for majoron DM and on its decay into two photons, to be discussed below.
It must be mentioned, however, that the existence of $U(1)_{B-L}$ breaking in the Lagrangian, as required for a non-zero majoron mass, can lead to severe fine-tuning issues. Some $U(1)$ breaking terms, such as $\sigma^3$ or even some Planck-scale suppressed operators, need to be heavily suppressed in order to keep the majoron mass small~\cite{Rothstein:1992rh}. Similar issues arise in axion models~\cite{Kallosh:1995hi}, but can often be solved by means of additional particles and symmetries.

The limit of interest in this article is the seesaw~\cite{Minkowski:1977sc} relation $m_D \ll \M_R$ in Eq.~\eqref{eq:neutrino_mass_matrix}, which leads to light neutrino masses of order $m_D^2/\M_R$, automatically suppressed with respect to the electroweak scale. This allows for a block-diagonalization and expansion in the small ratio $m_D/\M_R \sim \sqrt{d_l/d_h}$, leading to
\begin{align}
V &\simeq \matrixx{U^* & -\i U^* \sqrt{d_l} R^\dagger \sqrt{d_h^{-1}} \\
-\i \sqrt{d_h^{-1}} R \sqrt{d_l} & \mathbb{1}} ,\\
C &\simeq \matrixx{ \mathbb{1} & \i \sqrt{d_l} R^T \sqrt{d_h^{-1}} \\ 
-\i \sqrt{d_h^{-1}} R^* \sqrt{d_l} & 0 } ,\\
B &\simeq \matrixx{U & \quad\i U  \sqrt{d_l} R^T\sqrt{d_h^{-1}} } ,
\end{align}
where $d_l = \diag (m_1,m_2,m_3) \ll d_h = \diag (m_4,m_5,m_6)$, and $R = (R^T)^{-1}$ is a complex orthogonal $3\times 3$ matrix that arises in the Casas--Ibarra parametrization of $m_D = \i U \sqrt{d_l} R^T \sqrt{d_h}$ and describes the mixing between light and heavy neutrinos~\cite{Casas:2001sr}.
Since the mixing angles of the Pontecorvo--Maki--Nakagawa--Sakata matrix $U$ and the mass splittings $\Delta m_{21}^2$ and $|\Delta m_{32}^2|$ are known~\cite{Esteban:2016qun}, the free parameters in the seesaw limit are e.g.~$m_1$, $d_h$, $R$, $f$, $m_J$, and the three CP-violating phases in $U$.
It will prove useful to distinguish three different extreme hierarchies of light neutrinos:
\begin{align}
&\text{Normal Hierarchy (NH):} & &m_1\ll m_2 \ll m_3 \,,\\
&\text{Inverted Hierarchy (IH):} & &m_3\ll m_2 \simeq m_1\,,\\
&\text{Quasi-Degenerate (QD):} & &m_1\simeq m_2 \simeq m_3\,.
\end{align}
Assuming $m_{1,2,3}\ll m_J \ll m_{4,5,6}$, the majoron can decay into the light neutrinos, with partial widths proportional to $m_j^2$ due to the diagonal $J\nu\nu$ couplings in the seesaw limit:
\begin{align}
\Gamma &(J\to \nu\nu) \simeq \frac{m_J}{16\pi f^2}\sum_{j=1}^3 m_j^2 \label{eq:JtoNuNu}\\
&\simeq \frac{1}{\unit[3\times 10^{19}]{s}} \left(\frac{m_J}{\unit[1]{MeV}}\right)\left(\frac{\unit[10^{9}]{GeV}}{f}\right)^2\left(\frac{\sum_j m_j^2}{\unit[10^{-3}]{eV^2}}\right) .\nonumber
\end{align}
Neutrino oscillations give a lower bound on $\sum_j m_j^2$ of $\unit[2.6\times 10^{-3}]{eV^2}$ ($\unit[4.9\times 10^{-3}]{eV^2}$) for normal (inverted) mass ordering; Cosmology gives a conservative upper limit of $\unit[0.17]{eV^2}$~\cite{Aghanim:2016yuo}, to be used below for the QD regime, although much stronger limits even below $\unit[10^{-2}]{eV^2}$ are possible for certain combinations of datasets~\cite{Cuesta:2015iho,Aghanim:2016yuo,Giusarma:2016phn}.
We see that a majoron can easily be long-lived enough to be DM for typical seesaw scales, assuming $J\to\nu\nu$ to be the main decay channel.

\begin{figure}[t]
\includegraphics[width=0.48\textwidth]{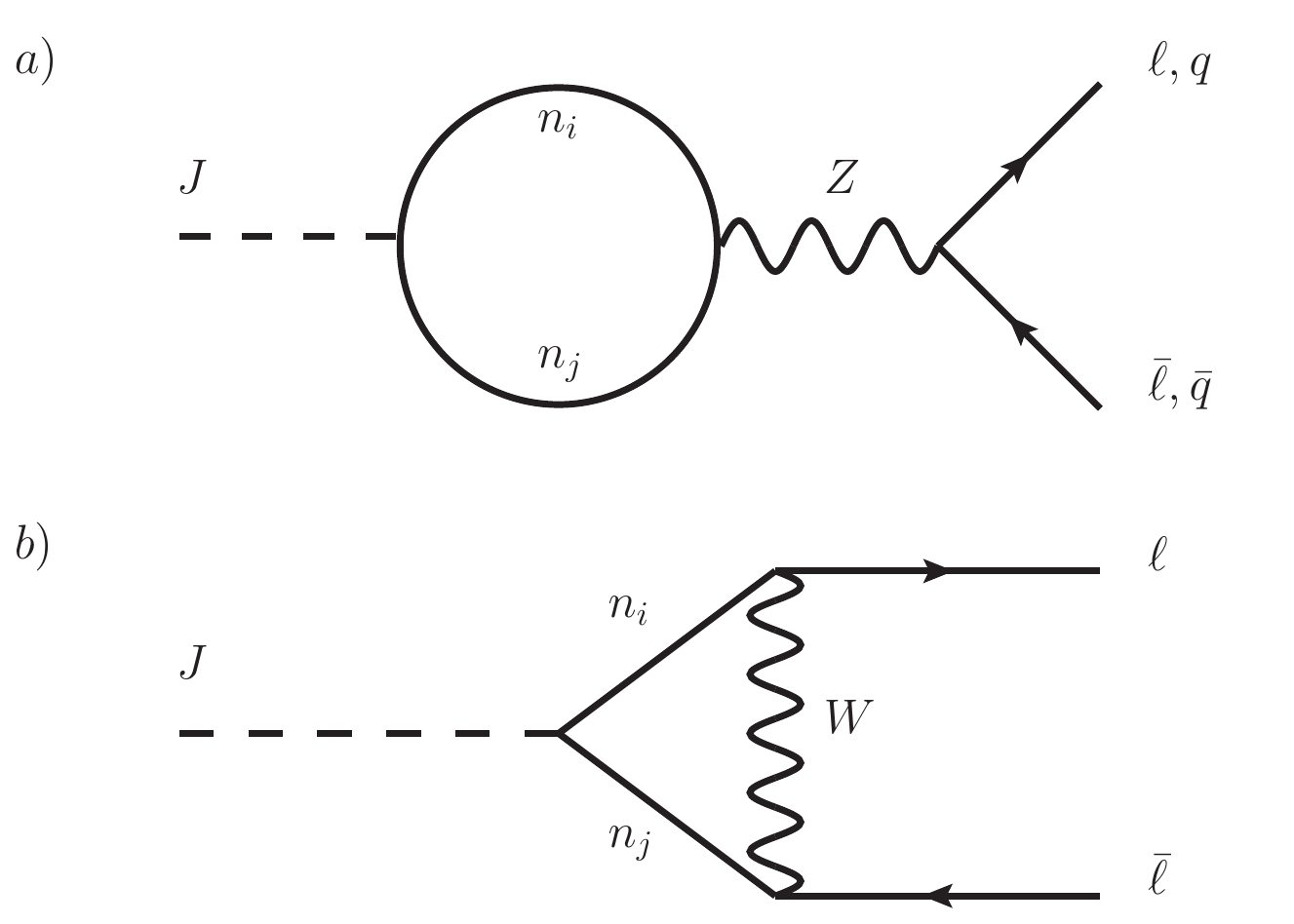}
\caption{
Loop-induced majoron couplings to charged fermions with the Majorana neutrino mass eigenstates $n_i$ running in the loops.}
\label{fig:Majoron_fermion_coupling}
\end{figure}

At the one-loop level one obtains a coupling of $J$ to charged fermions~\cite{Chikashige:1980ui,Pilaftsis:1993af}, which is crucial for majoron phenomenology. The Feynman diagrams are shown in Fig.~\ref{fig:Majoron_fermion_coupling} and give rise to the effective on-shell couplings 
\begin{align}
\L_J = \i J \bar{f}_1 (g_{Jf_1 f_2}^S+ g_{Jf_1 f_2}^P \gamma_5) f_2 \,,
\label{eq:effectiveJff}
\end{align}
with flavor-diagonal pseudoscalar quark couplings
\begin{align}
g_{J q q'}^P &\simeq \frac{m_q }{8\pi^2 v}\delta_{q q'} T^q_3 \, \tr K\,, & 
g_{J q q'}^S &= 0 \,,
\end{align}
and more involved lepton couplings,
\begin{align}
g_{J\ell{\ell'}}^P &\simeq \frac{m_\ell+m_{\ell'}}{16\pi^2 v} \left(\delta_{\ell \ell'} T^\ell_3\, \tr K +  K_{\ell\ell'} \right), & \\
g_{J\ell{\ell'}}^S &\simeq \frac{m_{\ell'}-m_\ell}{16\pi^2 v}  K_{\ell\ell'} \,,
\end{align}
to lowest order in the seesaw limit,
where $T^{d,\ell}_3 = -\tfrac12 = - T^u_3$.
The dimensionless hermitian $3\times 3$ matrix $K$ is defined as
\begin{align}
K \equiv  \frac{m_D m_D^\dagger}{v f}=\frac{1}{v f} U \sqrt{d_l} R^T d_h R^* \sqrt{d_l} U^\dagger .
\label{eq:K}
\end{align}
The partial width for the charged-fermion modes $J\to \bar f f$ is then given by
\begin{align}
\Gamma (J\to \bar q q) &\simeq \frac{3}{8\pi} |g_{J q q}^P|^2 m_J \,,\\
\Gamma (J\to \bar \ell \ell')  &\simeq \frac{1}{8\pi} \left(|g_{J \ell \ell'}^P|^2+|g_{J \ell \ell'}^S|^2 \right) m_J \,,
\end{align}
working again in the limit of small fermion masses.
A couple of remarks are in order:
\begin{itemize}
	\item All couplings $g_{J f_1 f_2}$ are proportional to the corresponding fermion masses as required for derivative couplings of Goldstone bosons. This in turn implies that the processes $J\to \overline{f_1} f_2$ are helicity suppressed as expected for a neutral spin-zero particle decaying into SM fermions. 
	\item The \emph{diagonal} fermion couplings $g_{J f f}$ are of pure pseudoscalar nature~\cite{Pilaftsis:1993af}.
	\item The \emph{off-diagonal} lepton couplings $g_{J \ell\ell'}$ are approximately chiral due to the hierarchy of charged-lepton masses,
	\begin{align}
	\qquad \L_{J\ell{\ell'}} \simeq -\frac{\i m_\ell}{8\pi^2 v} K_{\ell {\ell'}}\, J \,\bar{\ell} P_L {\ell'} + \hc ,
	\label{eq:offdiagonal}
	\end{align}
	for $m_\ell \gg m_{\ell'}$.
	\item The matrix $K$ is positive semi-definite if the lightest neutrino mass is zero and positive definite otherwise, with determinant
	\begin{align}
	\det K = \frac{1}{v^3 f^3}\prod_{j=1}^6 m_j \geq 0
	\end{align}
	and non-negative trace, $\tr K \geq 3 (\det K)^{1/3}$. All diagonal entries $K_{\ell\ell}$ are real and non-negative. Since $m_\text{lightest}=0$ is unstable under renormalization group evolution~\cite{Davidson:2006tg}, we can take $K$ to be strictly positive definite, which gives Schwarz inequalities on the off-diagonal entries,
	\begin{align}
|K_{\ell\ell'}|\leq \sqrt{K_{\ell\ell} K_{\ell'\ell'}} \leq \tr K \,.
\label{eq:inequality}
\end{align}
As a result, constraints on $\tr K$, e.g.~from $J\to \bar q q$, constrain \emph{all} entries of $K$, courtesy of its positive-definite nature.
\item 
From the definition $K = m_D m_D^\dagger/(v f)$ we can estimate a simple perturbativity condition by demanding $m_D/v < \sqrt{4\pi}$ (see also Ref.~\cite{Casas:2010wm}),
\begin{align}
|K_{\ell\ell'}| < \frac{4\pi v}{f} \simeq 3\times 10^{-6}\left( \frac{\unit[10^9]{GeV}}{f}\right) .
\label{eq:perturbative_unitarity}
\end{align} 
\emph{Typical} values for $K$ -- without fine-tuned matrix cancellations, i.e.~imaginary $R$ -- can on the other hand be estimated as 
\begin{align}
K\sim \frac{d_h d_l }{ f v}\sim \lambda \frac{d_l}{v}	\sim 2\times 10^{-13} \lambda\,,
\label{eq:typical}
\end{align}
 with the Yukawa coupling $\lambda$ from Eq.~\eqref{eq:lagrangian}. This is of course nothing but the result one obtains for one fermion generation, as calculated in Ref.~\cite{Chikashige:1980ui}.
\item As shown in Ref.~\cite{Davidson:2001zk}, the matrix $m_D m_D^\dagger$ (or $K$ in our case)  can be used to replace  $R$ and $d_h$ in the seesaw parametrization. In other words, the entire seesaw matrix from Eq.~\eqref{eq:neutrino_mass_matrix} can be reconstructed using low-energy neutrino parameters ($d_l$ and $U$) as well as $K$ and $f$. This is the parametrization of choice in this article, seeing as $K$ describes the physical couplings of the majoron to charged fermions and furthermore fulfills a number of useful inequalities that would be tedious to translate to e.g.~$R$.
It is quite remarkable that the seemingly lost high-energy seesaw parameters encoded in $m_D m_D^\dagger$ become available in the form of majoron couplings, allowing in principle to reconstruct the seesaw mechanism with low-energy data.
\end{itemize}

Due to the proportionality $\Gamma (J\to \bar f f)\propto m_f^2$, the dominant decay channel of $J$ is typically into the heaviest kinematically available fermion, but there are some notable loopholes: 1) the decay rates into charged fermions all scale with $K^2$ and can be made small compared to $J\to\nu\nu$ in the limit $\lambda \simeq \sqrt{2} d_h/f \ll 1$; 2) the diagonal lepton couplings $J\ell\ell$ are proportional to $\tr K - 2 K_{\ell\ell}$, which could be highly suppressed for up to two leptons despite $K$ being large~\cite{Pilaftsis:1993af,Pilaftsis:2008qt}. For example, the pattern $K_{ee} = K_{\mu\mu}\gg K_{\tau\tau}\simeq 0$ turns off the majoron couplings to $ee$ and $\mu\mu$. 

Since we are interested in majoron masses in the MeV--GeV range, the decays $J\to \bar uu, \bar dd, \bar ss, \bar cc$ should be replaced by appropriate decays into hadrons, which in particular moves the kinematic threshold from $m_J \simeq 2 m_u$ to $m_\pi$, with first allowed channel $J\to \pi \gamma\gamma$~\cite{Hiller:2004ii}, albeit heavily suppressed. Note that $J$ decays into pairs of pseudoscalar mesons are forbidden by CP, so the next threshold is $3 m_\pi$~\cite{McKeen:2008gd}. Seeing as not even the hadronic decay modes of a CP-\emph{even} Higgs-like scalar with mass between $0.1$--$\unit[10]{GeV}$ have been agreed-on in the literature (see e.g.~Ref.~\cite{Clarke:2013aya}), we will not attempt to derive the $J\to$\,hadron decay rates here, but leave them for future work. Estimates for a pseudoscalar's decay into three mesons can be found in Ref.~\cite{Dolan:2014ska}, assuming Higgs-like couplings. In the majoron model we have instead a Higgs-like coupling with additional sign-flip for up- and down quarks, just like in two-Higgs-doublet models of type I and X. The only hadronic decay used in the following is $J\to \bar b b$, which can be calculated reliably and will provide the best constraints on $K$ for $m_J\gtrsim \unit[10]{GeV}$.

Let us continue our discussion of majoron decay modes.
Still at the one-loop level one has virtual internal Bremsstrahlung, $J\to \bar f f \gamma$, simply by attaching photons to the diagrams in Fig.~\ref{fig:Majoron_fermion_coupling}. 
For quarks this merely gives the well-known final-state radiation spectrum, but the additional diagram with a $W$ boson gives a more interesting result for leptons.
The extra photon removes  the helicity suppression of the amplitude and leads to a  photon spectrum similar to the $s$-wave Majorana DM annihilation into $\bar f f\gamma$~\cite{Bergstrom:1989jr}, with characteristic shape for sizable photon energy
\begin{align}
\frac{1}{\Gamma (J\to \bar\ell \ell' \gamma)}\frac{\dd\Gamma (J\to \bar\ell \ell' \gamma)}{\dd x} \simeq 20 x^3 (1-x)\,,
\label{eq:VIBfeature}
\end{align}
for $x= 2 E_\gamma/m_J \in [0,1 ]$. In our case, the helicity suppression of the amplitude $\mathcal{A}\propto m_\ell$ is however replaced by an additional heavy-neutrino propagator, $\mathcal{A}\propto e \,m_J^3/d_h^2$, so the rate is of higher order in the seesaw expansion and hence strongly suppressed.
Bremsstrahlung will therefore not give testable signatures and will not be discussed further.

\begin{figure}[t]
\includegraphics[width=0.43\textwidth]{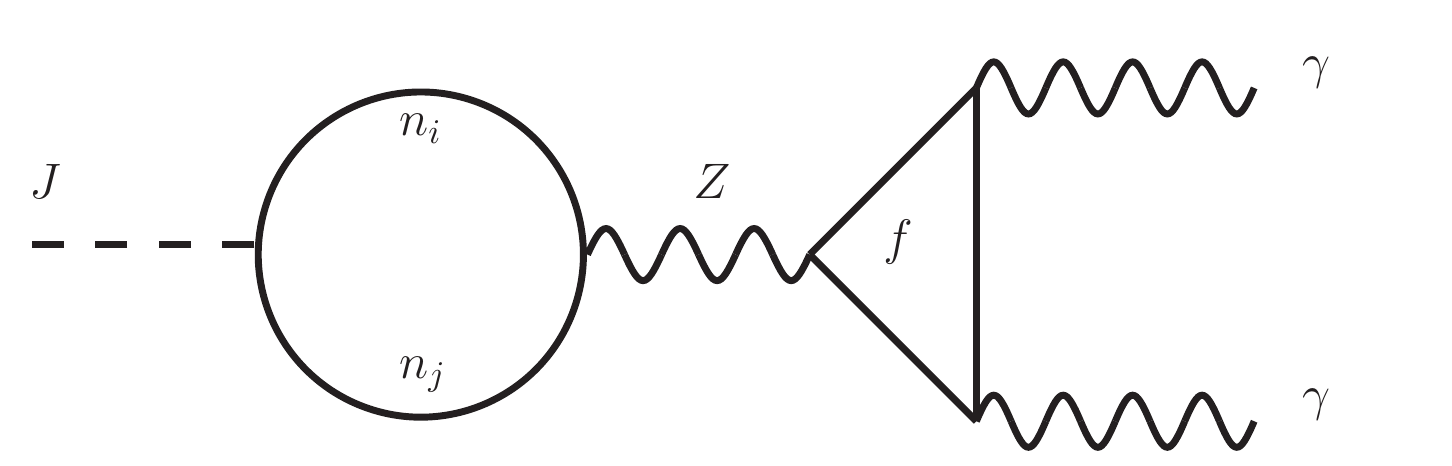}
\caption{
One two-loop contribution to $J\to\gamma\gamma$ via charged fermions $f$.}
\label{fig:Majoron_diphoton}
\end{figure}

Lastly, let us mention the possible decay mode $J\to \gamma\gamma$, which could be the prime discovery channel and has been discussed extensively in the literature for other models. For a massless majoron, the coupling to photons vanishes because the global $U(1)_{B-L}$ symmetry is anomaly free~\cite{Pilaftsis:1993af,Frigerio:2011in}. The coupling for a \emph{pseudo}-Goldstone boson then depends on the UV completion of the theory, i.e.~the details of how $m_J$ is generated (and whether the singlet has some admixture of a triplet majoron~\cite{Schechter:1981cv}). In the absence of $U(1)_{B-L}$-anomaly-inducing heavy fermions, our \emph{singlet}-majoron coupling to photons will be generated first at two loops. One contribution comes from the majoron mixing with the longitudinal component of the $Z$ boson, which then decays into two photons, see Fig.~\ref{fig:Majoron_diphoton}. Notice that only  fermion loops contribute to this piece of the amplitude, because similar diagrams with the $W$ boson and its Faddeev--Popov ghosts cancel each other~\cite{Garcia-Cely:2016hsk}.  The additional diagrams that arise from closing the leptonic lines in the $W$-boson loop of Fig.~\ref{fig:Majoron_fermion_coupling}\,b) are much more complicated to calculate, but we expect them to be further suppressed by the $W$ mass or even the heavy neutrino mass, so we will neglect them for now. Notice that such a separation of the diagrams is gauge invariant, as the corresponding amplitudes satisfy the Ward identities separately.  Note also that the $Z$-boson contribution depends on different parameters than the $W$-boson part (e.g.~quark masses), so it is not possible for the neglected diagrams to cancel the entire amplitude; a partial destructive interference could, of course, be possible.
Focusing only on the gauge-invariant part of the amplitude induced by $J$--$Z$ mixing, i.e.~Fig.~\ref{fig:Majoron_diphoton}, the two-loop rate takes the simple form
\begin{align}
\Gamma (J\to\gamma\gamma) \simeq \frac{\alpha^2 \left(\tr K\right)^2}{4096\pi^7} \frac{m_J^3}{v^2}  \left| \sum_f  N_c^f T_3^f Q_f^2 \,g \left(\frac{m_J^2}{4 m_f^2}\right)\right|^2 ,
\label{eq:diphoton}
\end{align}
with the color factor $N_c^q = 3$, $N_c^\ell =1$ and the loop function
\begin{align}
\begin{split}
g(x) &\equiv -\frac{1}{4 x} \left(\log [1-2 x + 2\sqrt{x (x-1)}]\right)^2\\
& = 1+ \frac{x}{3} + \frac{8x^2}{45} + \frac{4 x^3}{35}+\mathcal{O}(x^4)\,.
\end{split}
\end{align}
For $m_J \ll m_e$, the fermion-mass independent contributions cancel due to anomaly freedom, leading to a rate that is dominated by the lightest fermion,
\begin{align}
\Gamma (J\to\gamma\gamma) \simeq \frac{\alpha^2 \left(\tr K\right)^2}{1536^2 \pi^7} \frac{m_J^7}{v^2 m_e^4}\,,  \text{ for } m_J \ll m_e\,.
\end{align}
In particular, the coupling $J\gamma\gamma$ vanishes for $m_J=0$ as expected.
Up to a prefactor, the rate of Eq.~\eqref{eq:diphoton} is equivalent to the singlet--triplet majoron case, where the majoron--$Z$ mixing is induced already at tree level by the vacuum expectation value of an $SU(2)_L$ triplet $\Delta\to v_T/\sqrt{2}$~\cite{Bazzocchi:2008fh}. The singlet--triplet-majoron rate then follows from Eq.~\eqref{eq:diphoton} via $\tr K \to 32 \pi^2 v_T^2/(f v)$.

We stress once more that the above diphoton rate was obtained by considering only a (gauge-invariant) subset of two-loop diagrams. While we expect the remaining diagrams to be suppressed by $m_W$ or $d_h$ or even cancel completely, a full calculation is beyond the scope of this article. Furthermore, the rate can be modified by the details of the scalar (admixture of triplets or CP-violating mixing with the Higgs) and fermion sector ($B-L$ anomalous fermions that create a $J\gamma\gamma$ coupling at one-loop).
The reader should therefore be careful when interpreting the diphoton rate used here.

\section{Dark Matter}
\label{sec:DM}

Possible production mechanisms for majoron(-like) DM have been extensively discussed in Ref.~\cite{Frigerio:2011in}, assuming a restricted set of couplings in order to obtain predictions. For example, taking 
\begin{align}
\L_{\slashed{L}} &=\lambda_h \sigma^2 H^\dagger H +\hc\nonumber\\
  &\supset -\lambda_h J^2 H^\dagger H = -\tfrac12 m_J^2 J^2 (1+ h/v)^2
\end{align}
to be the only explicit $U(1)$ breaking term in the scalar potential and neglecting the $U(1)$-invariant portal $|\sigma|^2 H^\dagger H$, the relic density $\Omega_J$ of $J$ is completely fixed for a given mass $m_J$ (assuming small Yukawa couplings $\lambda$ to the heavy neutrinos). For sufficiently large $\lambda_h = m_J^2/v^2$, a thermal population of majorons is produced in the Early Universe from annihilations and the (inverse) decays of the Higgs boson; after the Higgs disappears from the thermal plasma, the DM density eventually freezes out.
The required value for $\lambda_h$ in this scenario is typically incompatible with constraints from direct detection or $h\to$\,invisible, at least in the mass range of interest here~\cite{Cline:2013gha}.
Another possible situation is to assume that the number of DM particles was negligible with respect to those of the SM after reheating. If the portal interaction coupling $\lambda_h$ has small values, the population of majorons never reaches thermal equilibrium; for temperatures much smaller than the Higgs mass  -- when the majoron decouples from the SM plasma -- its abundance approaches a constant value. This leads to~\cite{Hall:2009bx}
\begin{align}
 \Omega_J h^2 \simeq \frac{ 2.19 \times 10^{27} }{g^s_*\sqrt{g^\rho_*}} \frac{m_J\Gamma (h\to JJ)}{m_h^2},
\label{eq:Omega}
\end{align}
where $g^s_*$ and $g^\rho_*$ are the number of degrees of freedom contributing to the entropy and energy density when the majoron decouples. This is the \emph{freeze-in} mechanism, which obviously requires $m_J < m_h/2$ and a very small decay rate of the Higgs boson into majorons (automatically satisfying LHC constraints on $h\to$\,invisible). From the observed DM density, and taking $g^s_* \sim g^\rho_* \sim 100$, we obtain  $m_J \simeq \unit[2.7]{MeV}$ for the $\lambda_h = m_J^2/v^2$ case described above~\cite{Frigerio:2011in}. 

In a more general case, one can consider separate $U(1)$ breaking terms for the majoron mass and the Higgs portal, disentangling relic density and DM mass. For the freeze-out production mechanism, this is just the singlet DM scenario, heavily constrained and only viable around the Higgs resonance~\cite{Cline:2013gha}. For the production via freeze-in, Eq.~\eqref{eq:Omega} leads to  
\begin{align}
m_J \simeq  \left(\frac{\lambda_h}{2.0\times 10^{-10}}\right)^{-2} \unit{MeV}.
\label{eq:MJvsOmega}
\end{align}
Freeze-in is thus a viable  mechanism to produce majoron DM in the MeV and GeV range.  Other production mechanisms exist, see e.g.~Ref.~\cite{Frigerio:2011in} and references therein. Nevertheless, from now on we will remain agnostic about how DM was produced in the Early Universe and only assume that (cold) majorons constitute all the DM and that its mass lies below the electroweak scale. 
In any case, the specific indirect detection signatures discussed below do not depend on the details of the majoron mass generation or its production mechanism.

\subsubsection{Neutrino signatures}
\label{sec:neutrinos}

\begin{figure}[t]
\includegraphics[width=0.49\textwidth]{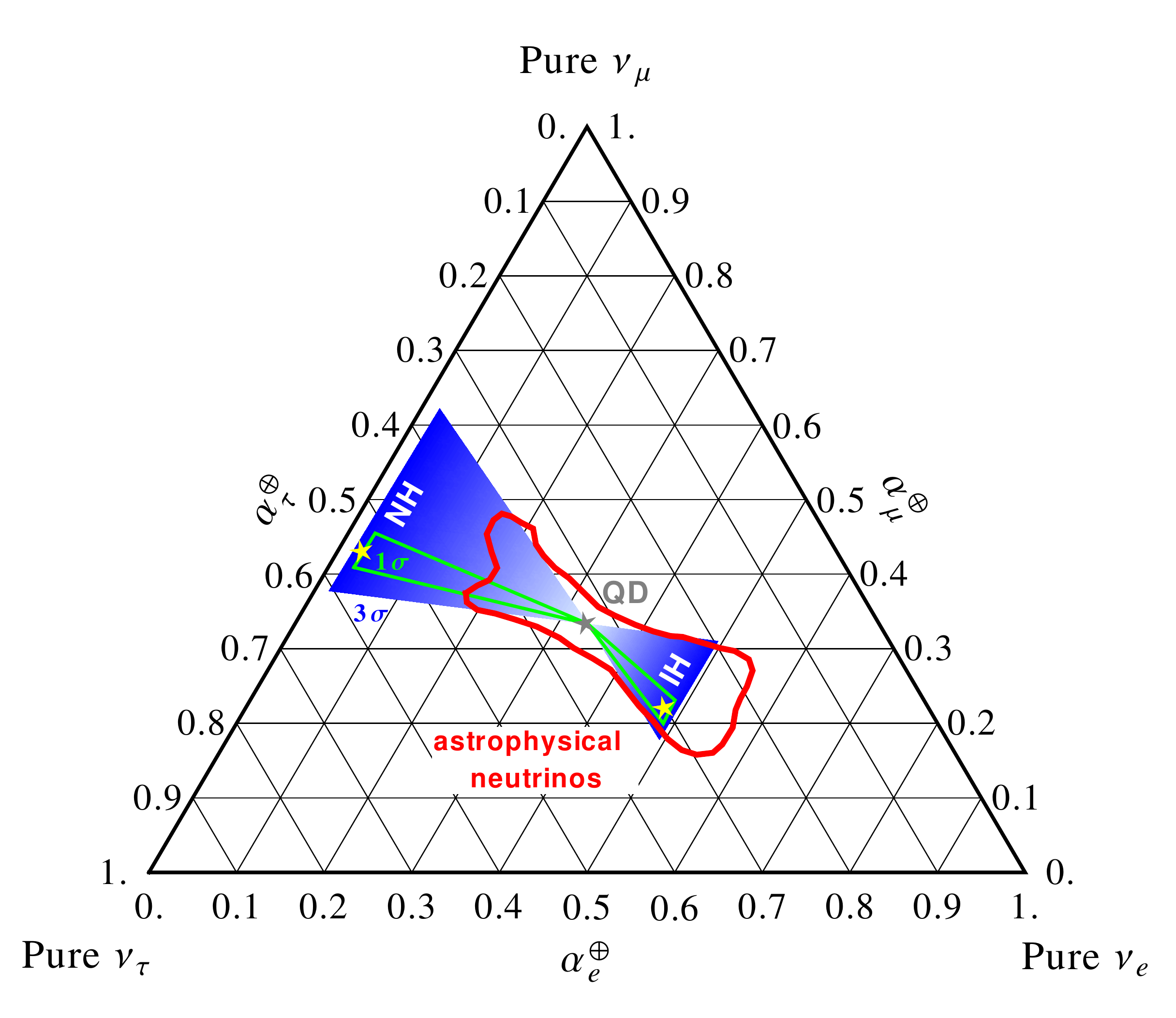}
\caption{
Majoron DM decay $J \to \nu\nu$ yields neutrino flavor ratios $\alpha_e :\alpha_\mu : \alpha_\tau$ that depend on the neutrino mass hierarchy.
The 1$\sigma$ (3$\sigma$) ranges of the neutrino-oscillation parameters from Ref.~\cite{Esteban:2016qun} correspond to the green (blue) lines; lighter colors correspond to a larger lightest-neutrino mass, converging to $1:1:1$ for the QD spectrum. 
The three stars denote the benchmark values of Eq.~\eqref{eq:benchmark_values}.
The expected flavor ratios from realistic astrophysical processes (e.g.~pion decay followed by averaged-out neutrino oscillations) fall in the red contour, taking into account the uncertainties in the mixing parameters ($95\%$~C.L.)~\cite{Vincent:2016nut}.
}
\label{fig:triangle}
\end{figure}

The only tree-level decay mode of the singlet majoron $J$ is into neutrinos, Eq.~\eqref{eq:JtoNuNu}, completely specified in terms of neutrino masses and $U(1)$ breaking scale $f$. An interesting side effect of the majoron coupling to neutrino \emph{mass eigenstates} is that the emitted neutrinos will not oscillate, resulting in flavor ratios that can be completely different from astrophysical sources~\cite{Bustamante:2015waa}; for a detailed discussion using the density-matrix formalism, see Appendix~\ref{sec:apA}. The branching ratio of $J$ decaying into $\nu_j$ is proportional to $m_j^2$, and $\nu_j$ contains a fraction $|U_{\ell j}|^2$ of flavor $\ell$, so the flavor composition of the majoron-decay neutrino flux is given by $\alpha_e :\alpha_\mu : \alpha_\tau$ with
\begin{align}
\alpha_\ell \equiv \frac{\sum_{j=1}^3 m_j^2 |U_{\ell j}|^2}{\sum_{j=1}^3 m_j^2} ,
\end{align}
normalized so that $\sum_\ell \alpha_\ell = 1$.
The self-conjugate Majorana nature ensures that $\alpha_{\ell} = \alpha_{\bar{\ell}}$.
Contrary to most other neutrino fluxes, these ratios are the same at the source, where DM decays, and on Earth, so $\alpha_\ell = \alpha_{\ell}^S= \alpha_{\ell}^\oplus$, up to matter effects inside the Earth.
See Fig.~\ref{fig:triangle} for an illustration using a ternary plot with a scan over the 1$\sigma$ and 3$\sigma$ ranges of the oscillation parameters obtained in Ref.~\cite{Esteban:2016qun}. 

The mixing angles $\theta_{23} \simeq \pi/4 \gg \theta_{13}$ result in an almost $\mu$--$\tau$-symmetric mixing matrix, i.e.~$|U_{\mu j}|\simeq |U_{\tau j}|$, which ensures $\alpha_\mu \simeq \alpha_\tau$ independent of the mass ordering. $\alpha_e$ on the other hand depends strongly on the neutrino mass hierarchy, with lowest value for NH ($\alpha_e \simeq \sin^2\theta_{13}$) and largest value for IH ($\alpha_e \simeq 1/2$). Using the best-fit values from Ref.~\cite{Esteban:2016qun} for the mixing angles, we obtain the following benchmark values for the flavor ratios in the hierarchical regime,
\begin{align}
&  & &\alpha_e :\alpha_\mu : \alpha_\tau \nonumber\\
&\text{NH:} & &0.03:0.43:0.54 \,,\nonumber\\
&\text{IH:} & &0.48:0.22:0.30\,,\label{eq:benchmark_values}\\
&\text{QD:} & &1:1:1\,,\nonumber
\end{align}
denoted by stars in Fig.~\ref{fig:triangle}. These are the values we will use in the following, but most results can be rescaled without much effort.
The NH composition with its tiny $\nu_e$ fraction $\alpha_e\simeq\sin^2\theta_{13}$ is particularly interesting, because there is no astrophysical mechanism that would suppress $\nu_e$ to such a degree without physics beyond the SM~\cite{Bustamante:2015waa}. This is illustrated in Fig.~\ref{fig:triangle}, where we also show the expected flavor ratios from astrophysical processes (red contour) under the assumption that the neutrino oscillations have been averaged out when the flux arrives at Earth~\cite{Bustamante:2015waa,Vincent:2016nut}, see Appendix~\ref{sec:apA} for details.
As can be seen, the NH region lies outside of the typical astrophysical expectation, making flavor ratios a potential discriminatory tool for DM detection.

\begin{figure*}[t]
\includegraphics[width=0.65\textwidth]{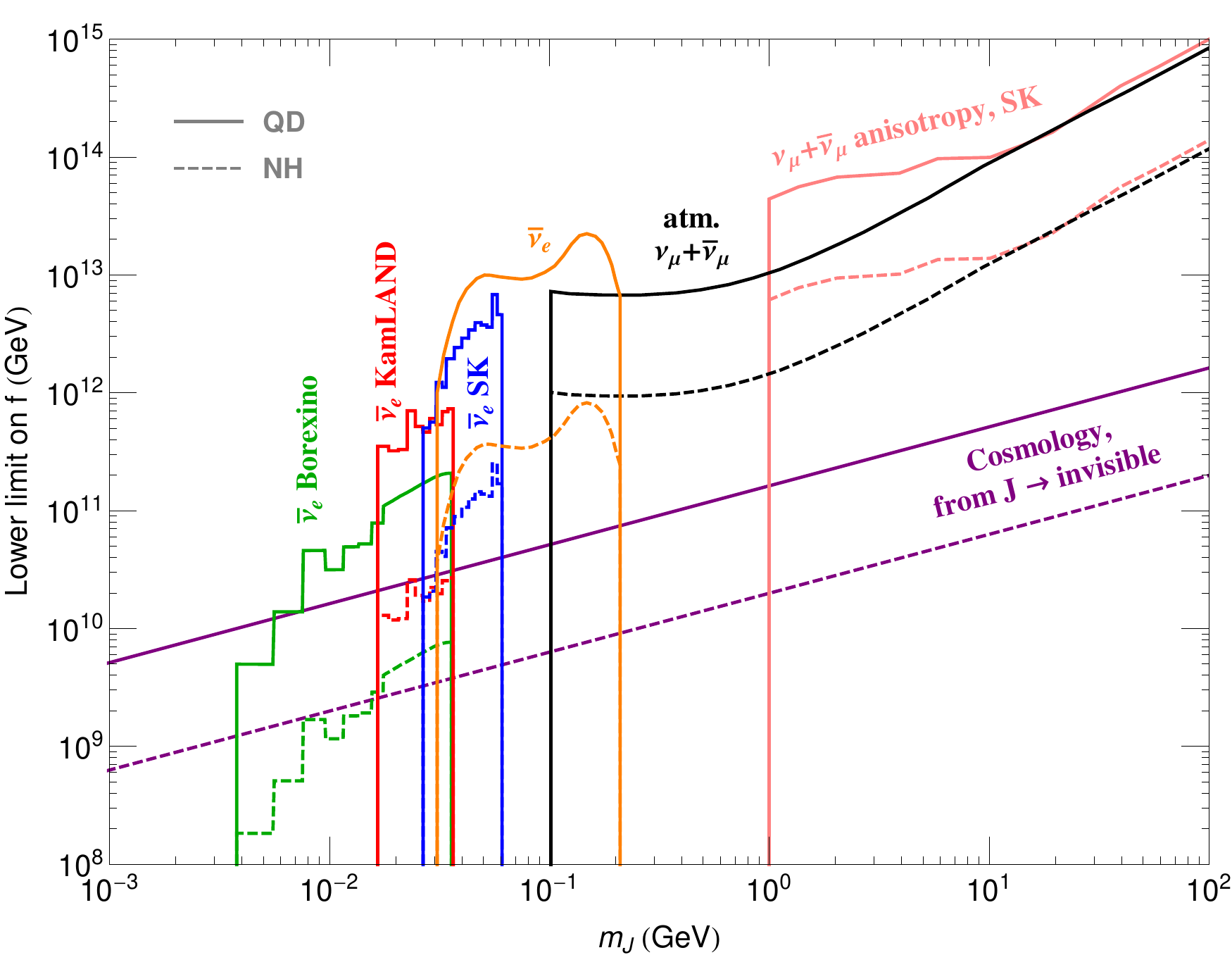}
\caption{
Lower bound on the $U(1)_{B-L}$ breaking scale $f$ for majoron $J$ DM, assuming QD (solid lines) or NH (dashed), IH lying in between.
The purple exclusion comes from cosmological constraints such as the CMB~\cite{Poulin:2016nat}.
Adopted limits from supernova $\bar\nu_e$ searches come from Borexino~\cite{Bellini:2010gn} (green), KamLAND~\cite{Collaboration:2011jza} (red), SK~\cite{Gando:2002ub,Zhang:2013tua} (blue), and reinterpreted SK data (orange)~\cite{PalomaresRuiz:2007ry}.
The black lines for $m_J >\unit[0.1]{GeV}$ come from a comparison with atmospheric $\nu_\mu$ spectra~\cite{PalomaresRuiz:2007ry}, while pink shows the preliminary limit from a designated DM search using angular-anisotropy SK data~\cite{FrankiewiczonbehalfoftheSuper-KamiokandeCollaboration:2016pkv}.
}
\label{fig:neutrino_lines}
\end{figure*}

Seeing as the majoron itself forms cold DM in our model, the neutrino spectrum with its line-like feature will be a much better discovery tool than the flavor ratios of Fig.~\ref{fig:triangle}. Let us mention, however, that the monochromatic signature becomes less important as soon as we consider $m_J$ above the electroweak scale; since the coupling to neutrinos of Eq.~\eqref{eq:Jnunu} also induces a coupling to the SM Higgs of the form $J\nu_j\nu_j (m_j/f)	 (1+h/v)^2$, the decay modes $J\to \nu\nu h(h)$ open up for $m_J > (2) m_h$, and in fact dominate over $J\to\nu\nu$ for $m_J\gtrsim\unit[10]{TeV}$~\cite{Dudas:2014bca}. The neutrino spectrum from $J\to \nu\nu h(h)$ is then obviously no longer monochromatic, but the flavor ratios of the primary neutrinos illustrated in Fig.~\ref{fig:triangle} continue to be valid. In addition, there will be secondary neutrinos with a different spectrum and flavor ratio from the $h$ decay and electroweak Bremsstrahlung. A thorough discussion of these effects will be discussed elsewhere, but we expect the flavor ratios of the secondary neutrinos to fall into the red contour of Fig.~\ref{fig:triangle}, because they are created as flavor eigenstates (see Appendix~\ref{sec:apA}).
Let us also mention that in models with a larger dark sector it is possible to obtain, for example, \emph{boosted} majorons that decay into a continuous neutrino spectrum, for which the flavor ratios could again be more important.

As mentioned above, the spectral feature of $J\to\nu\nu$ should serve as a sufficient discriminant from the continuous background. As shown in Ref.~\cite{Dudas:2014bca}, this two-body decay mode is suppressed compared to $J\to \nu\nu h(h)$ for $m_J\gtrsim\unit[10]{TeV}$, which induces a continuous spectrum. We will further restrict ourselves to masses $m_J < \unit[100]{GeV}$ in this analysis, in order to avoid discussing effects from e.g.~$J\to WW, ZZ$ that could be induced in some UV-completions of our model. We stress, however, that $J\to\nu\nu$ could still be an important discovery channel for majoron masses up to 10\,TeV, for which IceCube becomes the ideal observatory~\cite{Abbasi:2011eq,Aisati:2015vma}.
The neutrino (plus antineutrino) flux per flavor $\ell$ from the $J\to\nu\nu$ decay in our galaxy is given by~\cite{PalomaresRuiz:2007ry,Ibarra:2013zia}
\begin{align}
\frac{\dd \Phi_\ell}{\dd E_\nu} =\frac{{\cal J_{}}}{4 \pi} \frac{\alpha_\ell \Gamma (J\to\nu\nu)}{m_J} \frac{\dd N}{\dd E_\nu} \,,
\label{eq:nuflux}
\end{align}
where ${\cal J} = \int^\infty_0 \rho_\text{Halo} \dd s $ is the astrophysical factor associated to the DM density $\rho_\text{Halo}$  in the Milky Way  halo. For simplicity we write here the flux associated to the full sky, the general case for an angular signal is a straightforward generalization of this case. The ${\cal J}$-factor  introduces  uncertainties in the determination of the flux  because the precise shape of  $\rho_\text{Halo}$  is unknown in the center of the Galaxy. Nevertheless, in contrast to DM annihilations for which the ${\cal J}$ factor scales quadratically with $\rho_\text{Halo}$ and thus varies by many orders of magnitude depending on the assumptions on the DM halo, the uncertainty for DM decays is of less than one order of magnitude~\cite{PalomaresRuiz:2007ry} and the determination of neutrino fluxes or limits on them is more robust.  Notice that here we are neglecting the neutrino flux arising from DM decays outside our Galaxy, whose spectrum is in any case red-shifted and not necessarily line-like~\cite{Covi:2008jy,Covi:2009xn}.  

For the two-body decay $J\to\nu\nu$ we have $\dd N/\dd E_\nu = 2\delta (E_\nu - m_J/2)$, which is smeared out by the velocity distribution and detector resolution. 
Low-threshold neutrino detectors such as Borexino~\cite{Bellini:2010gn}, KamLAND~\cite{Collaboration:2011jza}, and SK~\cite{Gando:2002ub,Zhang:2013tua} give limits on the (monochromatic) flux of $\bar{\nu}_e$ from searches for the diffuse supernova neutrino background. Due to the large cross section and tagging possibilities, the detection channel of choice here is inverse beta decay $\bar{\nu}_e p \to n e^+$, which has a kinematic threshold of $E_\nu > \unit[1.8]{MeV}$. This makes it difficult to obtain limits for $m_J \lesssim\unit[4]{MeV}$, seeing as the background from reactor neutrinos also increases dramatically for such low energies. For $\unit[5]{MeV} \lesssim m_J < \unit[\mathcal{O}(100)]{MeV}$ on the other hand, searches for supernova $\bar{\nu}_e$ neutrinos give useful constraints on DM-induced neutrino fluxes, as can be seen in Fig.~\ref{fig:neutrino_lines}.
Note that in our notation this is a limit on the flux $\Phi_{\bar\nu_e} = \frac12 \Phi_e$, because only half of our electron neutrinos are antineutrinos. A near-future improvement of these limits is realistic, especially with the proposed Gadolinium-extension of SK~\cite{Fernandez:2015vhy}, which should reduce background and potentially reach the diffuse supernova regime. Even ton-scale liquid-xenon detectors build for the direct detection of DM, such as XENONnT, LZ or DARWIN, could be sensitive to $\mathcal{O}(\unit[10]{MeV})$ neutrino lines and might give useful information about the flavor ratios~\cite{Lang:2016zhv}.
In any case, dedicated DM searches by the experimental collaborations are desirable, especially considering the apparent gap of official limits between $m_J = \unit[60]{MeV}$ and GeV. Above GeV, we have preliminary SK limits on DM decay into muon neutrinos~\cite{FrankiewiczonbehalfoftheSuper-KamiokandeCollaboration:2016pkv}.
In the gap $\unit[60]{MeV}<m_J <\unit{GeV}$, we adapt the limits of Ref.~\cite{PalomaresRuiz:2007ry}, based on a reinterpretation of SK $\bar{\nu}_e$ data as well as a comparison to the well-understood atmospheric muon neutrino flux (see also Ref.~\cite{Covi:2009xn}). Here, we strongly urge the SK collaboration to check for neutrino lines, both in electron and muon neutrinos. Hyper-K is expected to further improve on the higher-energy region.

Depending on the neutrino mass hierarchy, these flux limits can be translated into a lower bound on the $U(1)_{B-L}$ breaking scale $f$, see Fig.~\ref{fig:neutrino_lines}. 
The latter is naturally strongest for QD, seeing as $\Gamma (J\to\nu\nu)\propto \sum_j m_j^2/f^2$ scales with the neutrino masses. In contrast, the weakest bounds arise for NH, which is quite obvious for limits that come from bounds on the total lifetime or from the $\alpha_e\simeq \sin^2\theta_{13}$ suppressed $\nu_e$ flux; surprisingly, limits from $\Phi_\mu$ lead to roughly the same bounds on $f$ for NH and IH, because of the accidental numerical relation
\begin{align}
\Phi_\mu \propto \sum_j m_j^2 |U_{\mu j}|^2 \simeq \begin{cases}\frac{m_3^2}{2} &\simeq \tfrac12 |\Delta m_{32}^2| \text{ for NH,} \\ \frac{m_1^2}{6} + \frac{m_2^2}{3} &\simeq\tfrac12 |\Delta m_{32}^2| \text{ for IH,}\end{cases}\nonumber
\end{align}
using tri-bimaximal mixing values as an approximation.
For the sake of clarity, it is therefore sufficient to discuss the limits for the regimes QD and NH in Fig.~\ref{fig:neutrino_lines}, as those associated to IH happen to fall in between.

Less direct limits on the $J\to\nu\nu$ decay come from cosmology. The most conservative bound is surely to demand $J$ to have a lifetime that exceeds the age of our Universe, $\tau \simeq\unit[4\times 10^{17}]{s}$. Better limits can be obtained by studying the effect that the decay of a non-relativistic DM particle into relativistic daughter particles has on e.g.~the matter power spectrum. A recent analysis provides a $95\%$~C.L.~constraint of order $\tau >\unit[5\times 10^{18}]{s}$~\cite{Audren:2014bca}. Future measurements of the cosmic microwave background (CMB), e.g.~by CORE, could improve the bound on $\tau$ by a factor of 2~\cite{DiValentino:2016foa}.
This is currently the only constraint on the $J\to\nu\nu$ mode for majoron masses below $\unit[4]{MeV}$ and will be hard to improve with line searches due to the huge neutrino background below $\unit[10]{MeV}$ from e.g.~reactor neutrinos~\cite{PalomaresRuiz:2007eu}.

\begin{figure*}[t]
\includegraphics[width=0.65\textwidth]{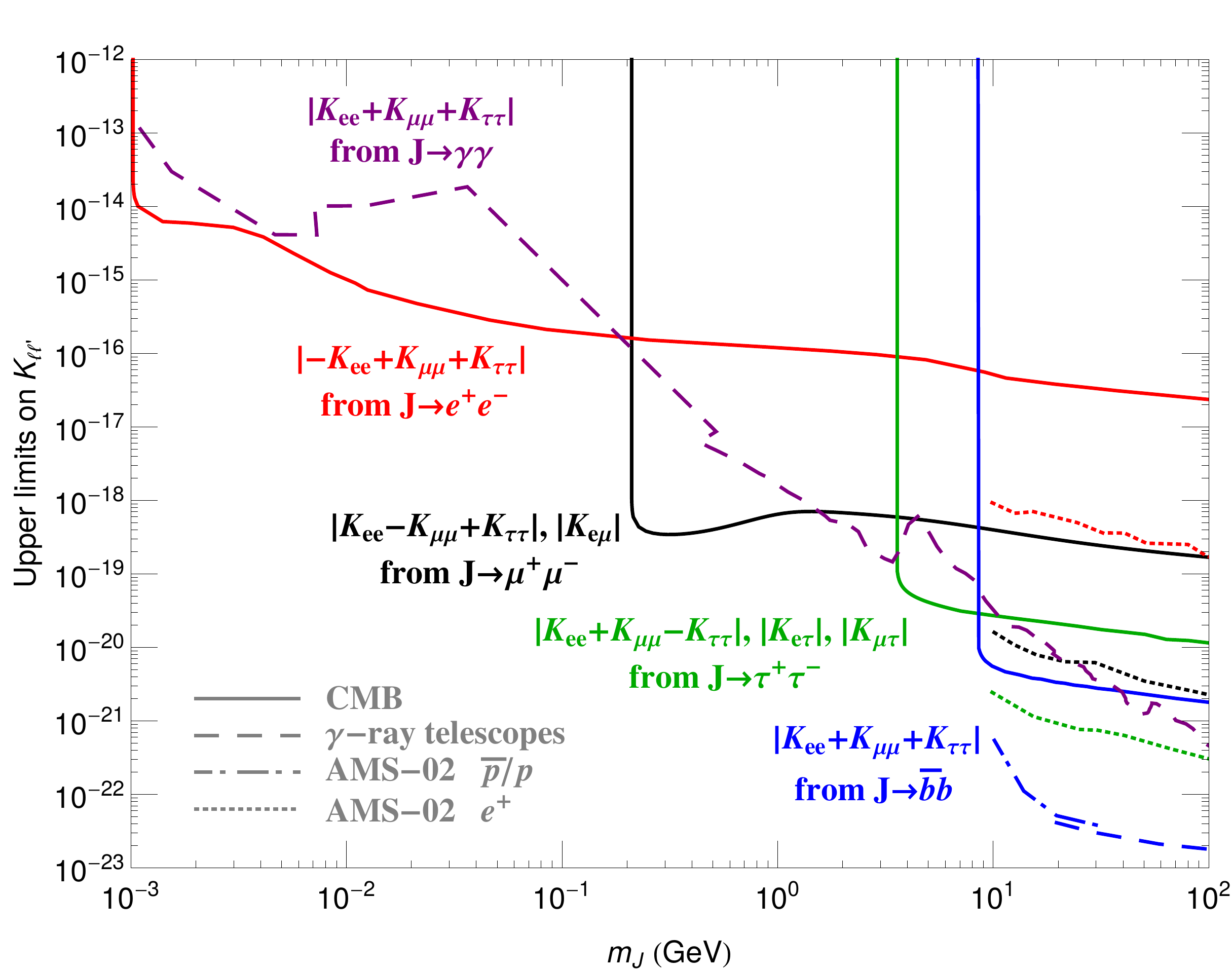}
\caption{
Upper bounds on the matrix elements $K_{\ell \ell'}$ or combinations of them from CMB measurements~\cite{Slatyer:2016qyl} and indirect DM searches with AMS-02~\cite{Ibarra:2013zia,Giesen:2015ufa}; $\gamma$-ray telescope limits on $J\to\gamma\gamma$ and $J\to \bar{b}b$ refer to INTEGRAL~\cite{Boyarsky:2007ge} for $m_J <\unit[7]{MeV}$, to  COMPTEL/EGRET~\cite{Yuksel:2007dr} for $\unit[7]{MeV}\leq m_J\leq \unit[400]{MeV}$, and to Fermi-LAT~\cite{Ackermann:2015lka,Cohen:2016uyg} for $m_J>\unit[400]{MeV}$.
For indirect DM searches, we only show the most constraining limits in a given channel. We remind the reader that $K$ is expected to have an order of magnitude of $10^{-13} \lambda$, where $\lambda$ is the Yukawa in Eq.~\eqref{eq:lagrangian}.
}
\label{fig:chargedfermions}
\end{figure*}

The limits on $f$ from Fig.~\ref{fig:neutrino_lines} can be translated into upper bounds on $|K_{\alpha\beta}|$ with the help of the perturbativity constraint of Eq.~\eqref{eq:perturbative_unitarity}. For $m_J = \unit[1]{MeV}$ ($\unit[100]{GeV}$) this implies $|K| < 5\times 10^{-6}$ ($3\times 10^{-11}$) for NH and about an order of magnitude stronger for QD. These limits are much weaker then the direct constraints from $J\to \bar f f'$ derived below (Fig.~\ref{fig:chargedfermions}), but are valid even if the $J$ decay is kinematically forbidden.

\subsubsection{Signatures from visible decay channels}
\label{sec:visible}

Having identified $\unit{MeV}\lesssim m_J \lesssim \unit[100]{GeV}$ as the region of interest where majoron DM can lead to a particularly clean observable flux of monochromatic neutrinos, let us discuss the constraints from the visible decay channels, i.e.~$J\to\bar f f$ at one loop and $J\to\gamma\gamma$ at two loop.
There are stringent constraints on DM decays into charged fermions from a wide
range of indirect searches, see e.g.~Ref.~\cite{Ibarra:2013cra} for a review.
In our model, the majoron decay modes into charged fermions all depend on the matrix $K= m_D m_D^\dagger/(v f)$ introduced in Eq.~\eqref{eq:K}. A crucial observation here is that $K$ does not depend on the low-energy neutrino parameters, but is a completely free parameter matrix in the seesaw limit, up to the inequalities given below Eq.~\eqref{eq:K}. Typical values can be estimated as $K\sim \frac{d_h d_l }{ f v} \simeq 2\times 10^{-13} \lambda$, but it is entirely possible to have values orders of magnitude larger or smaller. While the $J\to\nu\nu$ modes discussed above gave a direct limit on the seesaw scale $f$, the charged-fermion decay modes will give limits on the remaining parameters of our model, which are encoded in the elements of $K$.
The decays $J\to\nu\nu$ and $J\to\bar ff$ thus provide completely orthogonal information about majoron DM.

Majoron decays into charged leptons are in particular constrained by  the AMS-02 measurements of the positron flux in cosmic rays~\cite{Ibarra:2013zia}. The corresponding 95\%~C.L.~upper bounds on the $K$ matrix elements are shown in Fig.~\ref{fig:chargedfermions}. For masses above a few GeVs, other limits on the leptonic  decay channels arise from the diffuse-gamma-ray observations of the sky~\cite{ Cohen:2016uyg, Essig:2013goa, Cirelli:2012ut, Dugger:2010ys, Cirelli:2009dv, Zhang:2009ut, Yuksel:2007dr}, but these are typically less stringent than those of positrons for DM masses below 100 GeV.
In addition, for $m_J\gtrsim \unit[10]{GeV}$, the majoron decays dominantly into bottom quarks, which subsequently decay and fragment producing antiprotons.
The AMS-02 experiment has also measured the corresponding flux~\cite{Aguilar:2016kjl}, which, within astrophysical uncertainties, can be interpreted  as originating from only  cosmic ray collisions with the interstellar material~\cite{Giesen:2015ufa}. Slightly stronger bounds can be obtained with Fermi~\cite{Cohen:2016uyg}. This allows to set a strong upper bound on the decay rate into bottom quarks, $\tr K\lesssim 10^{-22}$ at 95\%~C.L.~for $m_J>\unit[10]{GeV}$, as shown in Fig.~\ref{fig:chargedfermions}.

The strongest of the indirect detection bounds is the one on $\tr K$ by  $J\to \bar b b$. As a matter of fact, this bound also applies to all entries of $K$ due to the inequality of Eq.~\eqref{eq:inequality}. Thus, majorons with masses greater than $\sim \unit[10]{GeV}$ are severely constrained, because such a small $K$ would require tiny Yukawa couplings $\lambda\sim 10^{-9}$. 
We expect constraints on $\tr K$ from the hadronic decay modes even below $\unit[10]{GeV}$, but the corresponding decays into mesons are difficult to calculate reliably.
Notice that because of these constraints, it is  hopeless to observe Majoron DM in direct  detection experiments looking for nuclear recoils. 

Below few GeVs, indirect detection bounds become very weak compared to CMB bounds. If DM decays into photons or charged particles during the time between recombination and reionization, when the Universe was transparent and no large-scale structures were formed, it injects energy into the photon--baryon fluid and potentially modifies the anisotropies of the CMB and its black-body shape. Consequently, the precise measurements of \textit{Planck}~\cite{Ade:2015xua} set stringent constraints on majoron decays into charged fermions. We calculate the corresponding  constraints\footnote{For second and third generation fermions, these limits were reported only for DM masses above \unit[10]{GeV}. Following the procedure described in Ref.~\cite{Slatyer:2016qyl}, we rederive the limits and extend them to lower masses.} on $K_{\ell \ell'}$ following Ref.~\cite{Slatyer:2016qyl}, and show them in Fig.~\ref{fig:chargedfermions}. These bounds are very important for two reasons. On the one hand, they constrain majoron decays at the MeV scale, where $J\to e^+e^-$ and $J\to \mu^+\mu^-$ are the dominant decay channels.  On the other hand, they do not suffer from astrophysical uncertainties such as those associated to halo DM densities or cosmic-ray propagation parameters.

Finally, let us discuss constraints from $J\to\gamma\gamma$, arguably the most popular decay channel for majorons~\cite{Bazzocchi:2008fh,Lattanzi:2013uza,Queiroz:2014yna}. Using our estimate for this two-loop decay of Eq.~\eqref{eq:diphoton}, we can translate $\gamma$-line limits from INTEGRAL~\cite{Boyarsky:2007ge}, COMPTEL/EGRET~\cite{Yuksel:2007dr}, and Fermi-LAT~\cite{Ackermann:2015lka,Cohen:2016uyg} into upper bounds on $\tr K$ (Fig.~\ref{fig:chargedfermions}). These $\gamma$-ray telescope limits are stronger than the corresponding CMB limits on $J\to\gamma\gamma$~\cite{Slatyer:2016qyl}, so we will not show them here. Due to the suppression by $\alpha^2$ and an additional loop compared to $J\to\bar f f$, the limits from $J\to\gamma\gamma$ are for the most part weaker than those from charged fermions. Nevertheless, the diphoton decays probe $\tr K$, which in turn limits all entries of $K$ via Eq.~\eqref{eq:inequality}, whereas the $J\to\ell \ell'$ decays only probe specific linear combination of $K$ elements. This makes the $J\to\gamma\gamma$ (and $J\to \bar bb$) constraints particularly interesting.
Future prospects for this channel are good, with considerable current effort to improve limits in the \emph{MeV gap} between $\unit[7]{MeV}\lesssim m_J\lesssim \unit[400]{MeV}$, for example by AdEPT~\cite{Boddy:2015efa} and e-ASTROGAM~\cite{Knodlseder:2016pey}. An improvement by several orders of magnitude seems feasible, which could open the door to a double-line observation in the MeV range, both in neutrinos and $\gamma$-rays.
For $m_J <\unit{MeV}$, the diphoton decay is the only feasible DM detection channel, seeing as sub-MeV neutrinos are extremely difficult to detect, especially when it comes to their spectral shape and flavor.

In summary, the constraints on majoron DM from its visible decay channels provide information on the model that is complementary to the main decay mode $J\to\nu\nu$. In the region of interest for neutrino lines, $\unit{MeV}\lesssim m_J \lesssim \unit[100]{GeV}$, the constraints on the elements of $K$ range from $10^{-13}$ to $10^{-23}$, which translates into \emph{typical} values for the Yukawa coupling $\lambda$ of $1$ to $10^{-10}$ via Eq.~\eqref{eq:typical}. This should not be taken too literally in the three-generation framework, but can give some idea about the level of tuning necessary to evade the bounds. In particular, the region $m_J \gtrsim\unit[10]{GeV}$ could be regarded as less motivated, which is however a highly subjective statement.

\section{Other constraints}
\label{sec:lfv}

For $m_J > m_{f_1}+ m_{f_2}$, the best constraints on the majoron couplings $g_{J f_1 f_2}$ come from the decay $J\to f_1 f_2$ or $J\to\gamma\gamma$, as we have seen in Fig.~\ref{fig:chargedfermions}. Let us briefly discuss limits from the \emph{production} of $J$, e.g.~from $f_1\to f_2 J$, which gives limits on $g_{J f_1 f_2}$ for $m_J < m_{f_1}- m_{f_2}$.
For $m_J >\unit{MeV}$, all these constraints turn out to be weaker than the perturbativity bounds of Eq.~\eqref{eq:perturbative_unitarity} in connection with the limits on $f$ from Fig.~\ref{fig:neutrino_lines}, which imply that $|K|$ can be at most $5\times 10^{-6}$ for $m_J\simeq\unit{MeV}$. Even stronger bounds apply when considering the limits from $J\to\gamma\gamma$ (Fig.~\ref{fig:chargedfermions}).
 We nevertheless list the direct constraints below for completeness, and stress that they can be relevant for $m_J<\unit{MeV}$ or if $J$ makes up only a subcomponent of DM.

The off-diagonal majoron couplings are directly constrained by the lepton flavor violating (LFV) decays $\ell \to \ell' J$~\cite{Pilaftsis:1993af,Feng:1997tn,Hirsch:2009ee}, with strongest bound in the muon sector, $\Br (\mu \to e J)< 2.6\times 10^{-6}$~\cite{Jodidio:1986mz}, and $\Br (\tau \to \ell J)<\mathcal{O}(10^{-3})$~\cite{Albrecht:1995ht}. 
The strong $\mu\to e J$ bound of Ref.~\cite{Jodidio:1986mz} rests on the assumption of isotropic electron emission; in our case, however, the emission is maximally \emph{anisotropic}, see Eq.~\eqref{eq:offdiagonal}, with dominant emission of the left-handed electron in the direction opposite to the muon polarization. This also happens to be the region where the background from $\mu\to e \nu\nu$ is largest, diminishing the limit by an order of magnitude~\cite{Bayes:2014lxz} to $|K_{\mu e}|\lesssim 1\times 10^{-5}$ for $m_J\ll m_\mu$.
An almost identical limit has been obtained long ago by considering $\mu\to e J \gamma$ with a massless $J$, which does not depend on the chirality properties of the $J\mu e$ coupling, but is of course further suppressed by $\alpha$ and phase space~\cite{Goldman:1987hy}. We checked explicitly that the rate for $\mu\to e J \gamma$ in our model is well described by the effective off-diagonal coupling of Eq.~\eqref{eq:offdiagonal} followed by Bremsstrahlung, leading to the same differential distributions given in Refs.~\cite{Goldman:1987hy,Hirsch:2009ee}.\footnote{Note an unfortunate typo in Ref.~\cite{Goldman:1987hy}, where the double-differential distributions are given as a function of $x= 2 E_e/m_\mu$, when it is actually $2 E_\gamma/m_\mu$.} 
Since the Bremsstrahlung rate formally diverges for small photon energies and small electron--photon opening angle, the number of events crucially depends on the detector resolution. It would be interesting to see how current and future experiments such as MEG and Mu3e can improve on these 30-year-old limits with their modern detectors~\cite{Hirsch:2009ee}, but this will be discussed elsewhere.

For $m_{\ell'}, m_J \ll m_\ell$, the partial widths are simply
\begin{align}
\frac{\Gamma (\ell \to \ell' J)}{\Gamma (\ell \to \ell' \nu_\ell \bar\nu_{\ell'})} \simeq \frac{3}{16 \pi^2}\frac{|(m_D m_D^\dagger)_{\ell\ell'}|^2}{m_\ell^2 f^2 }  = \frac{3}{16 \pi^2}\frac{v^2}{m_\ell^2}  |K_{\ell\ell'}|^2 ,
\end{align}
which then translate to the bounds
\begin{align}
&|K_{\mu e}|\lesssim 1\times 10^{-5}\,, &\text{ for }\quad &m_J\ll m_\mu\,,\nonumber\\
&|K_{\tau e}|\lesssim 6\times 10^{-3}\,, &\text{ for }\quad &m_J\ll m_\tau\,,\label{eq:LFV}\\
&|K_{\tau \mu}|\lesssim 9\times 10^{-3}\,, &\text{ for }\quad &m_J\ll m_\tau\,,\nonumber
\end{align}
neglecting the dependence on the majoron mass for simplicity. Perturbativity plus $J\to\nu\nu$ limits give stronger limits, unless $m_J <\unit{MeV}$; $J\to\gamma\gamma$ even requires $m_J <\unit[10]{keV}$ for LFV to be observable, at least if $J$ makes up $100\%$ of DM. Since such low-mass DM is typically not cold, a dedicated analysis is necessary to evaluate its validity.

Additional LFV in the form of $\ell\to\ell' \gamma$ arises from the right-handed neutrinos, which is independent of the majoron or breaking scale, with the strongest bound coming from $\Br (\mu\to e\gamma) < 4.2\times 10^{-13}$~\cite{TheMEG:2016wtm}. In the seesaw limit, $m_{1,2,3}\ll m_W \ll m_{4,5,6}$, the partial widths take the form~\cite{Cheng:1980tp,Heeck:2012fw}
\begin{align}
\frac{\Gamma (\ell \to \ell' \gamma)}{\Gamma (\ell \to \ell' \nu_\ell \bar\nu_{\ell'})} &\simeq \frac{3\alpha}{8\pi} \left| \left( m_D d_h^{-2} m_D^\dagger\right)_{\ell\ell'}\right|^2 ,
\end{align}
which has a different matrix structure than $K$, making it difficult to directly compare limits. 
In principle one can calculate the above for a given $d_l$, $U$, $K$ and $f$~\cite{Davidson:2001zk}, but the expression will be far from illuminating.
Large rates typically require some fine-tuning, e.g.~large $\Im (R)$ in the Casas--Ibarra parametrization, or a symmetry-motivated structure in $m_D$~\cite{Heurtier:2016iac}.
Let us focus on the case of degenerate heavy neutrinos, $d_h = M \times\mathbb{1}$, for which the above is proportional to $|K_{\ell\ell'}|^2$, allowing us to directly compare the two LFV rates,
\begin{align}
\frac{\Gamma (\ell \to \ell'\gamma)}{\Gamma (\ell \to \ell' J)} \simeq 2\pi\alpha \frac{m_\ell^2 }{M^2} \frac{f^2}{M^2}\,.
\end{align}
The ratio is heavily suppressed for $M\sim f \gg m_\ell$, making the majoron final state the prime LFV channel despite its more difficult signature; the photon rate can dominate for small Yukawa coupling, $\lambda =\sqrt2 M/f\ll 1$, implying not-too-heavy right-handed neutrinos. Both channels should hence be searched for experimentally.

The \emph{diagonal} majoron couplings, i.e.~the diagonal $K$ entries, are constrained via the $J$ coupling to electrons and quarks. At low energies, we typically require the couplings to \emph{nucleons} $N = (p,n)^T$ instead of quarks, which can be estimated naively as $J\overline{N} \i\gamma_5 \sigma_3 N\, m_N \tr K/(16\pi^2 v)$.
The coupling to quarks and nucleons is of particular interest, because it depends on $\tr K$, which automatically limits all entries in $K$ due to Eq.~\eqref{eq:inequality}, even the LFV couplings. Limits on (light) pseudoscalars can be readily found in the literature, usually assuming an effective coupling to fermions that is then used to calculate scattering processes etc.; this will be at best an \emph{approximately} accurate procedure in our model, because our effective $Jff$ couplings from Eq.~\eqref{eq:effectiveJff} are by construction only valid for \emph{on-shell} particles. As such, scattering processes -- which necessarily involve off-shell particles -- would have to be calculated from scratch using the loop diagrams to obtain the correct dependence of the cross sections on our parameters.

A full calculation of all the required scattering rates being beyond the scope of this work, let us assume that the $Jff$ couplings provide a reasonable estimate for majoron scattering.
For $m_J<\unit[10]{keV}$, the best limits then come from astrophysics and imply
\begin{align}
|K_{ee}-K_{\mu\mu} - K_{\tau\tau}| < 2\times 10^{-5} \,, &&
\tr K < 10^{-5}\,,
\end{align}
from the electron~\cite{Raffelt:1994ry} and nucleon coupling~\cite{Raffelt:2012sp}, respectively.
 For $m_J$ up to $\unit[100]{keV}$ one has (slightly weaker) direct-detection bounds on $g_{J ee}^P$ from EDELWEISS~\cite{Armengaud:2013rta}, XENON~\cite{Aprile:2014eoa}, XMASS~\cite{Abe:2014zcd}, and MAJORANA~\cite{Abgrall:2016tnn}, assuming $J$ to be DM; this gives $|K_{ee}-K_{\mu\mu} - K_{\tau\tau}| \lesssim 10^{-4}$~\cite{Abe:2014zcd} for $m_J = \unit[100]{keV}$, roughly ten orders of magnitude weaker than the bound at $m_J =\unit[\mathcal{O}(1)]{MeV}$ (Fig.~\ref{fig:chargedfermions}). 
The couplings to quarks are much less constrained for $m_J >\unit[10]{keV}$; since there are no flavor-changing processes in the quark sector mediated by the majoron at the one-loop level, quark-flavor constraints are suppressed. Going to the next loop level we can estimate constraints from $K\to\pi J$ etc.~following Ref.~\cite{Dolan:2014ska}, which give constraints $\tr K \lesssim 2\times 10^{-2}$ for $m_J <\unit[100]{MeV}$, much weaker for larger $m_J$.
In the region of interest in this article, $\unit{MeV} \leq m_J \leq \unit[100]{GeV}$, majoron \emph{production} gives weaker limits on $K$ than perturbativity in combination with the neutrino limits on $f$, and much weaker than the $J\to\gamma\gamma$ bounds.

Lastly, let us mention another signature of our model: neutrinoless double beta decay $0\nu\beta\beta$~\cite{Rodejohann:2011mu}. In the seesaw limit, the amplitude for this $\Delta L = 2$ process is dominated by light-neutrino exchange, proportional to $(U d_l U^T)_{ee} = \sum_{j=1}^3 U_{e j}^2 m_j$. This is in particular sensitive to the Majorana CP phases in $U$, which cannot be measured via neutrino oscillations. Current experiments probe the QD regime, with limits of order $|(U d_l U^T)_{ee}| < \unit[0.2]{eV}$~\cite{Gando:2012zm}. Future experiments are expected to ultimately reach the IH regime, while NH leads to discouragingly small rates. The observation of $0\nu\beta\beta$ would be an incredible discovery and prove beyond doubt that neutrinos are Majorana particles, leading further credence to the seesaw mechanism. This would of course be good news for our majoron model at hand, as it would in particular fix the rather strong dependence of e.g.~$J\to\nu\nu$ on the neutrino hierarchy. It should be mentioned, however, that our (sub-MeV) majoron DM gives completely negligible rates for the associated $0\nu\beta\beta J$ process $(A,Z)\to (A,Z+2) + 2 e^- + J$~\cite{Rodejohann:2011mu}, seeing as the majoron couplings to neutrinos $m_\nu/f$ are minuscule. The discovery of such a mode would therefore strongly hint at a more complicated majoron realization.
Due to the small $J\nu\nu$ coupling, supernova constraints are also easily evaded~\cite{Heurtier:2016otg}.

\section{Conclusion}
\label{sec:conclusion}

In this work, we have revisited the singlet majoron model, in which lepton number is a nearly exact symmetry that is spontaneously broken at the seesaw scale. The corresponding pseudo-Goldstone boson, the majoron, is stable on cosmological scales due to its highly suppressed couplings and can act as DM.
One of the most remarkable features of this model is the prediction of monochromatic neutrinos arising from DM decays, practically testable at energies between MeV (e.g.~Borexino) and 100\,GeV (e.g.~Super-K), potentially up to 10\,TeV (IceCube). We urge the experimental collaborations to perform designated searches for such low-energy DM-induced neutrino lines.
Since the majoron couples directly to the neutrino mass eigenstates, the decay neutrinos do not oscillate and have flavor ratios on Earth that depend strongly on the neutrino mass hierarchy, see Figs.~\ref{fig:triangle} and \ref{fig:neutrino_lines}. In particular, the electron-neutrino flux is suppressed compared to the other flavors for the normal mass hierarchy.

Other constraints on the model arise from the majoron couplings to charged fermions, induced at the one-loop level, and the decay into two photons, induced by two-loop diagrams. We have provided a convenient and compact three-generation parametrization of these couplings in terms of the matrix $m_D m_D^\dagger$, which contains precisely those seesaw parameters that are usually unobservable at low energies. A measurement of the majoron couplings could then in principle complete our knowledge of the seesaw mechanism. In the DM context, majoron decays into charged fermions and diphotons are constrained by CMB observations and indirect DM searches, which put strong limits on $m_D m_D^\dagger$, especially for $m_J >\unit[10]{GeV}$, as illustrated in Fig.~\ref{fig:chargedfermions}. 
Our parametrization also allows us to study constraints from lepton flavor violation; the rates for anisotropic $\ell\to\ell' J$ turn out to be small for $m_J\gtrsim \unit{MeV}$ if $J$ makes up all of DM, but $\ell\to\ell' \gamma$ can be observable for not-too-heavy right-handed neutrinos.

\section*{Acknowledgements}
We would like to thank Anna Lamperstorfer, Thomas Hambye, Sergio Palomares-Ruiz and Hiren Patel for useful discussions. 
CGC is supported by the IISN and the Belgian Federal Science Policy through the Interuniversity Attraction Pole P7/37 ``Fundamental Interactions''.
JH is a postdoctoral researcher of the F.R.S.-FNRS.
We acknowledge the use of \texttt{Package-X}~\cite{Patel:2015tea,Patel:2016fam} and \texttt{JaxoDraw}~\cite{Binosi:2003yf}.

\appendix

\section{Neutrino Oscillations}
\label{sec:apA}

An astrophysical source producing neutrinos with an energy $E$ and  flavor ratios $\alpha_{\ell}^S$ leads to the density matrix $\rho^S_{\ell\ell'} = \alpha_{\ell}^S\delta_{\ell \ell'}$. Neutrinos oscillate during their travel from the source to Earth, as can be seen from the fact that $\rho^S$ does not commute with the propagation  Hamiltonian, given  in the neutrino-mass basis by  $H_{ij}\simeq (E+ \frac{m_i^2}{2E}) \delta_{ij}$. In fact, the density matrix describing the flux of neutrinos after a distance $L$ at Earth reads $\rho^\oplus = e^{-\i H L} \rho^S e^{\i H L}$, or more precisely,
\begin{equation}
\rho^\oplus_{\ell\ell'}\simeq \sum_{i,j,\ell''} U_{\ell i} \, e^{-\i \frac{m_i^2 L}{2E}}\, U^*_{\ell''i}\,\alpha_{\ell''}^S\, U_{\ell'' j} \,e^{\i \frac{m_j^2 L }{2E}}\,U^*_{l'j}\,.
\end{equation}
For a sufficiently large oscillation length $L$, neutrino oscillations average out and $\exp\{-\i \frac{m_i^2-m_j^2}{2E}L\}\to \delta_{ij}$, which leads to  $\rho^\oplus_{\ell\ell'} \simeq \sum_{i\,\ell''} U_{\ell i}  U^*_{\ell''i}\alpha_{\ell''}^S U_{\ell''i} \,U^*_{l'i}\,$. The flavor composition at Earth, given by the diagonal elements of the density matrix, is thus
\begin{align}
&\alpha^\oplus \simeq P \alpha^S  &\text{        with    }& &P_{\ell\ell'} = \sum_{i} |U_{\ell i}|^2  |U_{\ell'i}|^2 \,. 
\end{align}
By varying the oscillation angles within the ranges allowed by neutrino experiments and assuming an arbitrary composition of flavors at the source, we obtain the red contour of Fig.~\ref{fig:triangle}.

The situation is different for neutrinos arising from majoron decay. In this case, the 
branching ratios associated to  $J\to\nu_i\nu_j$ are proportional to $m_j^2 \delta_{ij}$, at least in the lowest seesaw order we consider. Accordingly, the density matrix at the source is diagonal in the mass basis and commutes with the Hamiltonian. As a result, $\rho^\oplus = e^{-\i H L} \rho^S e^{\i H L} = \rho^S$ and therefore $\alpha_{\ell}^\oplus = \alpha_{\ell}^S$.

\bibliographystyle{utcaps_mod}
\bibliography{BIB}

\providecommand{\href}[2]{#2}\begingroup\raggedright\begin{thebibliography}{100}

\bibitem{Minkowski:1977sc}
P.~Minkowski, ``{\em {$\mu \to e\gamma$ at a Rate of One Out of $10^{9}$ Muon
  Decays?}},''
\href{http://dx.doi.org/10.1016/0370-2693(77)90435-X}{Phys. Lett. {\normalfont
  \bfseries B67} (1977)  421--428}.

\bibitem{Fukugita:1986hr}
M.~Fukugita and T.~Yanagida, ``{\em {Baryogenesis Without Grand
  Unification}},''
\href{http://dx.doi.org/10.1016/0370-2693(86)91126-3}{Phys. Lett. {\normalfont
  \bfseries B174} (1986)  45--47}.

\bibitem{Chikashige:1980ui}
Y.~Chikashige, R.~N. Mohapatra, and R.~D. Peccei, ``{\em {Are There Real
  Goldstone Bosons Associated with Broken Lepton Number?}},''
\href{http://dx.doi.org/10.1016/0370-2693(81)90011-3}{Phys. Lett. {\normalfont
  \bfseries B98} (1981)  265--268}.

\bibitem{Schechter:1981cv}
J.~Schechter and J.~W.~F. Valle, ``{\em {Neutrino Decay and Spontaneous
  Violation of Lepton Number}},''
\href{http://dx.doi.org/10.1103/PhysRevD.25.774}{Phys. Rev. {\normalfont
  \bfseries D25} (1982)  774}.

\bibitem{Rothstein:1992rh}
I.~Z. Rothstein, K.~S. Babu, and D.~Seckel, ``{\em {Planck scale symmetry
  breaking and majoron physics}},''
  \href{http://dx.doi.org/10.1016/0550-3213(93)90368-Y}{Nucl. Phys.
  {\normalfont \bfseries B403} (1993)  725--748},
\href{http://arxiv.org/abs/hep-ph/9301213}{{\normalfont \ttfamily
  arXiv:hep-ph/9301213}}.

\bibitem{Berezinsky:1993fm}
V.~Berezinsky and J.~W.~F. Valle, ``{\em {The KeV majoron as a dark matter
  particle}},'' \href{http://dx.doi.org/10.1016/0370-2693(93)90140-D}{Phys.
  Lett. {\normalfont \bfseries B318} (1993)  360--366},
\href{http://arxiv.org/abs/hep-ph/9309214}{{\normalfont \ttfamily
  arXiv:hep-ph/9309214}}.

\bibitem{Lattanzi:2007ux}
M.~Lattanzi and J.~W.~F. Valle, ``{\em {Decaying warm dark matter and neutrino
  masses}},'' \href{http://dx.doi.org/10.1103/PhysRevLett.99.121301}{Phys. Rev.
  Lett. {\normalfont \bfseries 99} (2007)  121301},
\href{http://arxiv.org/abs/0705.2406}{{\normalfont \ttfamily arXiv:0705.2406}}.

\bibitem{Bazzocchi:2008fh}
F.~Bazzocchi, M.~Lattanzi, S.~Riemer-S{\o}rensen, and J.~W.~F. Valle, ``{\em
  {X-ray photons from late-decaying majoron dark matter}},''
  \href{http://dx.doi.org/10.1088/1475-7516/2008/08/013}{JCAP {\normalfont
  \bfseries 0808} (2008)  013},
\href{http://arxiv.org/abs/0805.2372}{{\normalfont \ttfamily arXiv:0805.2372}}.

\bibitem{Frigerio:2011in}
M.~Frigerio, T.~Hambye, and E.~Masso, ``{\em {Sub-GeV dark matter as
  pseudo-Goldstone from the seesaw scale}},''
  \href{http://dx.doi.org/10.1103/PhysRevX.1.021026}{Phys. Rev. {\normalfont
  \bfseries X1} (2011)  021026},
\href{http://arxiv.org/abs/1107.4564}{{\normalfont \ttfamily arXiv:1107.4564}}.

\bibitem{Lattanzi:2013uza}
M.~Lattanzi, S.~Riemer-S{\o}rensen, M.~T\'ortola, and J.~W.~F. Valle, ``{\em
  {Updated CMB and x- and $\gamma$-ray constraints on Majoron dark matter}},''
  \href{http://dx.doi.org/10.1103/PhysRevD.88.063528}{Phys. Rev. {\normalfont
  \bfseries D88} (2013)  063528},
\href{http://arxiv.org/abs/1303.4685}{{\normalfont \ttfamily arXiv:1303.4685}}.

\bibitem{Queiroz:2014yna}
F.~S. Queiroz and K.~Sinha, ``{\em {The Poker Face of the Majoron Dark Matter
  Model: LUX to keV Line}},''
  \href{http://dx.doi.org/10.1016/j.physletb.2014.06.016}{Phys. Lett.
  {\normalfont \bfseries B735} (2014)  69--74},
\href{http://arxiv.org/abs/1404.1400}{{\normalfont \ttfamily arXiv:1404.1400}}.

\bibitem{Wang:2016vfj}
W.~Wang and Z.-L. Han, ``{\em {Global $U(1)_{L}$ Breaking in Neutrinophilic
  2HDM: From LHC Signatures to X-Ray Line}},''
  \href{http://dx.doi.org/10.1103/PhysRevD.94.053015}{Phys. Rev. {\normalfont
  \bfseries D94} (2016)  053015},
\href{http://arxiv.org/abs/1605.00239}{{\normalfont \ttfamily
  arXiv:1605.00239}}.

\bibitem{Boucenna:2014uma}
S.~M. Boucenna, S.~Morisi, Q.~Shafi, and J.~W.~F. Valle, ``{\em {Inflation and
  majoron dark matter in the seesaw mechanism}},''
  \href{http://dx.doi.org/10.1103/PhysRevD.90.055023}{Phys. Rev. {\normalfont
  \bfseries D90} (2014)  055023},
\href{http://arxiv.org/abs/1404.3198}{{\normalfont \ttfamily arXiv:1404.3198}}.

\bibitem{Higaki:2014dwa}
T.~Higaki, R.~Kitano, and R.~Sato, ``{\em {Neutrinoful Universe}},''
  \href{http://dx.doi.org/10.1007/JHEP07(2014)044}{JHEP {\normalfont \bfseries
  07} (2014)  044},
\href{http://arxiv.org/abs/1405.0013}{{\normalfont \ttfamily arXiv:1405.0013}}.

\bibitem{Sierra:2014sta}
D.~Aristizabal~Sierra, M.~Tortola, J.~W.~F. Valle, and A.~Vicente, ``{\em
  {Leptogenesis with a dynamical seesaw scale}},''
  \href{http://dx.doi.org/10.1088/1475-7516/2014/07/052}{JCAP {\normalfont
  \bfseries 1407} (2014)  052},
\href{http://arxiv.org/abs/1405.4706}{{\normalfont \ttfamily arXiv:1405.4706}}.

\bibitem{Mohapatra:1982tc}
R.~N. Mohapatra and G.~Senjanovic, ``{\em {The Superlight Axion and Neutrino
  Masses}},''
\href{http://dx.doi.org/10.1007/BF01577819}{Z. Phys. {\normalfont \bfseries
  C17} (1983)  53--56}.

\bibitem{Langacker:1986rj}
P.~Langacker, R.~D. Peccei, and T.~Yanagida, ``{\em {Invisible Axions and Light
  Neutrinos: Are They Connected?}},''
\href{http://dx.doi.org/10.1142/S0217732386000683}{Mod. Phys. Lett.
  {\normalfont \bfseries A1} (1986)  541}.

\bibitem{Ballesteros:2016euj}
G.~Ballesteros, J.~Redondo, A.~Ringwald, and C.~Tamarit, ``{\em {Unifying
  inflation with the axion, dark matter, baryogenesis and the seesaw
  mechanism}},'' \href{http://dx.doi.org/10.1103/PhysRevLett.118.071802}{Phys.
  Rev. Lett. {\normalfont \bfseries 118} (2017)  071802},
\href{http://arxiv.org/abs/1608.05414}{{\normalfont \ttfamily
  arXiv:1608.05414}}.

\bibitem{Ballesteros:2016xej}
G.~Ballesteros, J.~Redondo, A.~Ringwald, and C.~Tamarit, ``{\em {Standard
  Model-Axion-Seesaw-Higgs Portal Inflation. Five problems of particle physics
  and cosmology solved in one stroke}},''
\href{http://arxiv.org/abs/1610.01639}{{\normalfont \ttfamily
  arXiv:1610.01639}}.

\bibitem{DiLuzio:2016sbl}
L.~Di~Luzio, F.~Mescia, and E.~Nardi, ``{\em {Redefining the Axion Window}},''
  \href{http://dx.doi.org/10.1103/PhysRevLett.118.031801}{Phys. Rev. Lett.
  {\normalfont \bfseries 118} (2017)  031801},
\href{http://arxiv.org/abs/1610.07593}{{\normalfont \ttfamily
  arXiv:1610.07593}}.

\bibitem{Lindner:2010rr}
M.~Lindner, A.~Merle, and V.~Niro, ``{\em {Enhancing Dark Matter Annihilation
  into Neutrinos}},'' \href{http://dx.doi.org/10.1103/PhysRevD.82.123529}{Phys.
  Rev. {\normalfont \bfseries D82} (2010)  123529},
\href{http://arxiv.org/abs/1005.3116}{{\normalfont \ttfamily arXiv:1005.3116}}.

\bibitem{Gustafsson:2013gca}
M.~Gustafsson, T.~Hambye, and T.~Scarna, ``{\em {Effective Theory of Dark
  Matter Decay into Monochromatic Photons and its Implications: Constraints
  from Associated Cosmic-Ray Emission}},''
  \href{http://dx.doi.org/10.1016/j.physletb.2013.06.032}{Phys. Lett.
  {\normalfont \bfseries B724} (2013)  288--295},
\href{http://arxiv.org/abs/1303.4423}{{\normalfont \ttfamily arXiv:1303.4423}}.

\bibitem{Aisati:2015ova}
C.~El~Aisati, M.~Gustafsson, T.~Hambye, and T.~Scarna, ``{\em {Dark Matter
  Decay to a Photon and a Neutrino: the Double Monochromatic Smoking Gun
  Scenario}},'' \href{http://dx.doi.org/10.1103/PhysRevD.93.043535}{Phys. Rev.
  {\normalfont \bfseries D93} (2016)  043535},
\href{http://arxiv.org/abs/1510.05008}{{\normalfont \ttfamily
  arXiv:1510.05008}}.

\bibitem{Arina:2015zoa}
C.~Arina, S.~Kulkarni, and J.~Silk, ``{\em {Monochromatic neutrino lines from
  sneutrino dark matter}},''
  \href{http://dx.doi.org/10.1103/PhysRevD.92.083519}{Phys. Rev. {\normalfont
  \bfseries D92} (2015)  083519},
\href{http://arxiv.org/abs/1506.08202}{{\normalfont \ttfamily
  arXiv:1506.08202}}.

\bibitem{Queiroz:2016zwd}
F.~S. Queiroz, C.~E. Yaguna, and C.~Weniger, ``{\em {Gamma-ray Limits on
  Neutrino Lines}},''
  \href{http://dx.doi.org/10.1088/1475-7516/2016/05/050}{JCAP {\normalfont
  \bfseries 1605} (2016)  050},
\href{http://arxiv.org/abs/1602.05966}{{\normalfont \ttfamily
  arXiv:1602.05966}}.

\bibitem{PalomaresRuiz:2007ry}
S.~Palomares-Ruiz, ``{\em {Model-Independent Bound on the Dark Matter
  Lifetime}},'' \href{http://dx.doi.org/10.1016/j.physletb.2008.05.040}{Phys.
  Lett. {\normalfont \bfseries B665} (2008)  50--53},
\href{http://arxiv.org/abs/0712.1937}{{\normalfont \ttfamily arXiv:0712.1937}}.

\bibitem{Beacom:2006tt}
J.~F. Beacom, N.~F. Bell, and G.~D. Mack, ``{\em {General Upper Bound on the
  Dark Matter Total Annihilation Cross Section}},''
  \href{http://dx.doi.org/10.1103/PhysRevLett.99.231301}{Phys. Rev. Lett.
  {\normalfont \bfseries 99} (2007)  231301},
\href{http://arxiv.org/abs/astro-ph/0608090}{{\normalfont \ttfamily
  arXiv:astro-ph/0608090}}.

\bibitem{Yuksel:2007ac}
H.~Yuksel, S.~Horiuchi, J.~F. Beacom, and S.~Ando, ``{\em {Neutrino Constraints
  on the Dark Matter Total Annihilation Cross Section}},''
  \href{http://dx.doi.org/10.1103/PhysRevD.76.123506}{Phys. Rev. {\normalfont
  \bfseries D76} (2007)  123506},
\href{http://arxiv.org/abs/0707.0196}{{\normalfont \ttfamily arXiv:0707.0196}}.

\bibitem{PalomaresRuiz:2007eu}
S.~Palomares-Ruiz and S.~Pascoli, ``{\em {Testing MeV dark matter with neutrino
  detectors}},'' \href{http://dx.doi.org/10.1103/PhysRevD.77.025025}{Phys. Rev.
  {\normalfont \bfseries D77} (2008)  025025},
\href{http://arxiv.org/abs/0710.5420}{{\normalfont \ttfamily arXiv:0710.5420}}.

\bibitem{Kachelriess:2007aj}
M.~Kachelriess and P.~D. Serpico, ``{\em {Model-independent dark matter
  annihilation bound from the diffuse $\gamma$ ray flux}},''
  \href{http://dx.doi.org/10.1103/PhysRevD.76.063516}{Phys. Rev. {\normalfont
  \bfseries D76} (2007)  063516},
\href{http://arxiv.org/abs/0707.0209}{{\normalfont \ttfamily arXiv:0707.0209}}.

\bibitem{Bell:2008ey}
N.~F. Bell, J.~B. Dent, T.~D. Jacques, and T.~J. Weiler, ``{\em {Electroweak
  Bremsstrahlung in Dark Matter Annihilation}},''
  \href{http://dx.doi.org/10.1103/PhysRevD.78.083540}{Phys. Rev. {\normalfont
  \bfseries D78} (2008)  083540},
\href{http://arxiv.org/abs/0805.3423}{{\normalfont \ttfamily arXiv:0805.3423}}.

\bibitem{Bellini:2010gn}
{\normalfont \bfseries Borexino}, G.~Bellini {\em et al.}, ``{\em {Study of
  solar and other unknown anti-neutrino fluxes with Borexino at LNGS}},''
  \href{http://dx.doi.org/10.1016/j.physletb.2010.12.030}{Phys. Lett.
  {\normalfont \bfseries B696} (2011)  191--196},
\href{http://arxiv.org/abs/1010.0029}{{\normalfont \ttfamily arXiv:1010.0029}}.

\bibitem{Collaboration:2011jza}
{\normalfont \bfseries KamLAND}, A.~Gando {\em et al.}, ``{\em {A study of
  extraterrestrial antineutrino sources with the KamLAND detector}},''
  \href{http://dx.doi.org/10.1088/0004-637X/745/2/193}{Astrophys. J.
  {\normalfont \bfseries 745} (2012)  193},
\href{http://arxiv.org/abs/1105.3516}{{\normalfont \ttfamily arXiv:1105.3516}}.

\bibitem{Gando:2002ub}
{\normalfont \bfseries Super-Kamiokande}, Y.~Gando {\em et al.}, ``{\em {Search
  for $\bar\nu_e$ from the sun at Super-Kamiokande I}},''
  \href{http://dx.doi.org/10.1103/PhysRevLett.90.171302}{Phys. Rev. Lett.
  {\normalfont \bfseries 90} (2003)  171302},
\href{http://arxiv.org/abs/hep-ex/0212067}{{\normalfont \ttfamily
  arXiv:hep-ex/0212067}}.

\bibitem{Zhang:2013tua}
{\normalfont \bfseries Super-Kamiokande}, H.~Zhang {\em et al.}, ``{\em
  {Supernova Relic Neutrino Search with Neutron Tagging at
  Super-Kamiokande-IV}},''
  \href{http://dx.doi.org/10.1016/j.astropartphys.2014.05.004}{Astropart. Phys.
  {\normalfont \bfseries 60} (2015)  41--46},
\href{http://arxiv.org/abs/1311.3738}{{\normalfont \ttfamily arXiv:1311.3738}}.

\bibitem{Heeck:2012fw}
J.~Heeck, ``{\em {Seesaw parametrization for $n$ right-handed neutrinos}},''
  \href{http://dx.doi.org/10.1103/PhysRevD.86.093023}{Phys. Rev. {\normalfont
  \bfseries D86} (2012)  093023},
\href{http://arxiv.org/abs/1207.5521}{{\normalfont \ttfamily arXiv:1207.5521}}.

\bibitem{Pilaftsis:1993af}
A.~Pilaftsis, ``{\em {Astrophysical and terrestrial constraints on singlet
  Majoron models}},'' \href{http://dx.doi.org/10.1103/PhysRevD.49.2398}{Phys.
  Rev. {\normalfont \bfseries D49} (1994)  2398--2404},
\href{http://arxiv.org/abs/hep-ph/9308258}{{\normalfont \ttfamily
  arXiv:hep-ph/9308258}}.

\bibitem{Akhmedov:1992hi}
E.~K. Akhmedov, Z.~G. Berezhiani, R.~N. Mohapatra, and G.~Senjanovic, ``{\em
  {Planck scale effects on the majoron}},''
  \href{http://dx.doi.org/10.1016/0370-2693(93)90887-N}{Phys. Lett.
  {\normalfont \bfseries B299} (1993)  90--93},
\href{http://arxiv.org/abs/hep-ph/9209285}{{\normalfont \ttfamily
  arXiv:hep-ph/9209285}}.

\bibitem{Gu:2010ys}
P.-H. Gu, E.~Ma, and U.~Sarkar, ``{\em {Pseudo-Majoron as Dark Matter}},''
  \href{http://dx.doi.org/10.1016/j.physletb.2010.05.012}{Phys. Lett.
  {\normalfont \bfseries B690} (2010)  145--148},
\href{http://arxiv.org/abs/1004.1919}{{\normalfont \ttfamily arXiv:1004.1919}}.

\bibitem{Kallosh:1995hi}
R.~Kallosh, A.~D. Linde, D.~A. Linde, and L.~Susskind, ``{\em {Gravity and
  global symmetries}},'' \href{http://dx.doi.org/10.1103/PhysRevD.52.912}{Phys.
  Rev. {\normalfont \bfseries D52} (1995)  912--935},
\href{http://arxiv.org/abs/hep-th/9502069}{{\normalfont \ttfamily
  arXiv:hep-th/9502069}}.

\bibitem{Casas:2001sr}
J.~A. Casas and A.~Ibarra, ``{\em {Oscillating neutrinos and $\mu\to e
  \gamma$}},'' \href{http://dx.doi.org/10.1016/S0550-3213(01)00475-8}{Nucl.
  Phys. {\normalfont \bfseries B618} (2001)  171--204},
\href{http://arxiv.org/abs/hep-ph/0103065}{{\normalfont \ttfamily
  arXiv:hep-ph/0103065}}.

\bibitem{Esteban:2016qun}
I.~Esteban, M.~C. Gonzalez-Garcia, M.~Maltoni, I.~Martinez-Soler, and
  T.~Schwetz, ``{\em {Updated fit to three neutrino mixing: exploring the
  accelerator-reactor complementarity}},''
  \href{http://dx.doi.org/10.1007/JHEP01(2017)087}{JHEP {\normalfont \bfseries
  01} (2017)  087}, \href{http://arxiv.org/abs/1611.01514}{{\normalfont
  \ttfamily arXiv:1611.01514}}.
NuFit 3.0 from \url{http://www.nu-fit.org}.

\bibitem{Aghanim:2016yuo}
{\normalfont \bfseries Planck}, N.~Aghanim {\em et al.}, ``{\em {Planck
  intermediate results. XLVI. Reduction of large-scale systematic effects in
  HFI polarization maps and estimation of the reionization optical depth}},''
  \href{http://dx.doi.org/10.1051/0004-6361/201628890}{Astron. Astrophys.
  {\normalfont \bfseries 596} (2016)  A107},
\href{http://arxiv.org/abs/1605.02985}{{\normalfont \ttfamily
  arXiv:1605.02985}}.

\bibitem{Cuesta:2015iho}
A.~J. Cuesta, V.~Niro, and L.~Verde, ``{\em {Neutrino mass limits: robust
  information from the power spectrum of galaxy surveys}},''
  \href{http://dx.doi.org/10.1016/j.dark.2016.04.005}{Phys. Dark Univ.
  {\normalfont \bfseries 13} (2016)  77--86},
\href{http://arxiv.org/abs/1511.05983}{{\normalfont \ttfamily
  arXiv:1511.05983}}.

\bibitem{Giusarma:2016phn}
E.~Giusarma, M.~Gerbino, O.~Mena, S.~Vagnozzi, S.~Ho, and K.~Freese, ``{\em
  {Improvement of cosmological neutrino mass bounds}},''
  \href{http://dx.doi.org/10.1103/PhysRevD.94.083522}{Phys. Rev. {\normalfont
  \bfseries D94} (2016)  083522},
\href{http://arxiv.org/abs/1605.04320}{{\normalfont \ttfamily
  arXiv:1605.04320}}.

\bibitem{Davidson:2006tg}
S.~Davidson, G.~Isidori, and A.~Strumia, ``{\em {The smallest neutrino
  mass}},'' \href{http://dx.doi.org/10.1016/j.physletb.2007.01.015}{Phys. Lett.
  {\normalfont \bfseries B646} (2007)  100--104},
\href{http://arxiv.org/abs/hep-ph/0611389}{{\normalfont \ttfamily
  arXiv:hep-ph/0611389}}.

\bibitem{Casas:2010wm}
J.~A. Casas, J.~M. Moreno, N.~Rius, R.~Ruiz~de Austri, and B.~Zaldivar, ``{\em
  {Fair scans of the seesaw. Consequences for predictions on LFV processes}},''
  \href{http://dx.doi.org/10.1007/JHEP03(2011)034}{JHEP {\normalfont \bfseries
  03} (2011)  034},
\href{http://arxiv.org/abs/1010.5751}{{\normalfont \ttfamily arXiv:1010.5751}}.

\bibitem{Davidson:2001zk}
S.~Davidson and A.~Ibarra, ``{\em {Determining seesaw parameters from weak
  scale measurements?}},''
  \href{http://dx.doi.org/10.1088/1126-6708/2001/09/013}{JHEP {\normalfont
  \bfseries 09} (2001)  013},
\href{http://arxiv.org/abs/hep-ph/0104076}{{\normalfont \ttfamily
  arXiv:hep-ph/0104076}}.

\bibitem{Pilaftsis:2008qt}
A.~Pilaftsis, ``{\em {Electroweak Resonant Leptogenesis in the Singlet Majoron
  Model}},'' \href{http://dx.doi.org/10.1103/PhysRevD.78.013008}{Phys. Rev.
  {\normalfont \bfseries D78} (2008)  013008},
\href{http://arxiv.org/abs/0805.1677}{{\normalfont \ttfamily arXiv:0805.1677}}.

\bibitem{Hiller:2004ii}
G.~Hiller, ``{\em {B physics signals of the lightest CP odd Higgs in the NMSSM
  at large tan beta}},''
  \href{http://dx.doi.org/10.1103/PhysRevD.70.034018}{Phys. Rev. {\normalfont
  \bfseries D70} (2004)  034018},
\href{http://arxiv.org/abs/hep-ph/0404220}{{\normalfont \ttfamily
  arXiv:hep-ph/0404220}}.

\bibitem{McKeen:2008gd}
D.~McKeen, ``{\em {Constraining Light Bosons with Radiative $\Upsilon (1S)$
  Decays}},'' \href{http://dx.doi.org/10.1103/PhysRevD.79.015007}{Phys. Rev.
  {\normalfont \bfseries D79} (2009)  015007},
\href{http://arxiv.org/abs/0809.4787}{{\normalfont \ttfamily arXiv:0809.4787}}.

\bibitem{Clarke:2013aya}
J.~D. Clarke, R.~Foot, and R.~R. Volkas, ``{\em {Phenomenology of a very light
  scalar (100 MeV $< m_h <$ 10 GeV) mixing with the SM Higgs}},''
  \href{http://dx.doi.org/10.1007/JHEP02(2014)123}{JHEP {\normalfont \bfseries
  02} (2014)  123},
\href{http://arxiv.org/abs/1310.8042}{{\normalfont \ttfamily arXiv:1310.8042}}.

\bibitem{Dolan:2014ska}
M.~J. Dolan, F.~Kahlhoefer, C.~McCabe, and K.~Schmidt-Hoberg, ``{\em {A taste
  of dark matter: Flavour constraints on pseudoscalar mediators}},''
  \href{http://dx.doi.org/10.1007/JHEP03(2015)171}{JHEP {\normalfont \bfseries
  03} (2015)  171}, \href{http://arxiv.org/abs/1412.5174}{{\normalfont
  \ttfamily arXiv:1412.5174}}.
[Erratum: JHEP07,103(2015)].

\bibitem{Bergstrom:1989jr}
L.~Bergstrom, ``{\em {Radiative Processes in Dark Matter Photino
  Annihilation}},''
\href{http://dx.doi.org/10.1016/0370-2693(89)90585-6}{Phys. Lett. {\normalfont
  \bfseries B225} (1989)  372--380}.

\bibitem{Garcia-Cely:2016hsk}
C.~Garcia-Cely and A.~Rivera, ``{\em {General calculation of the cross section
  for dark matter annihilations into two photons}},''
  \href{http://dx.doi.org/10.1088/1475-7516/2017/03/054}{JCAP {\normalfont
  \bfseries 1703} (2017)  054},
\href{http://arxiv.org/abs/1611.08029}{{\normalfont \ttfamily
  arXiv:1611.08029}}.

\bibitem{Cline:2013gha}
J.~M. Cline, K.~Kainulainen, P.~Scott, and C.~Weniger, ``{\em {Update on scalar
  singlet dark matter}},''
  \href{http://dx.doi.org/10.1103/PhysRevD.88.055025}{Phys. Rev. {\normalfont
  \bfseries D88} (2013)  055025},
  \href{http://arxiv.org/abs/1306.4710}{{\normalfont \ttfamily
  arXiv:1306.4710}}.
[Erratum: Phys. Rev.D92,no.3,039906(2015)].

\bibitem{Hall:2009bx}
L.~J. Hall, K.~Jedamzik, J.~March-Russell, and S.~M. West, ``{\em {Freeze-In
  Production of FIMP Dark Matter}},''
  \href{http://dx.doi.org/10.1007/JHEP03(2010)080}{JHEP {\normalfont \bfseries
  03} (2010)  080},
\href{http://arxiv.org/abs/0911.1120}{{\normalfont \ttfamily arXiv:0911.1120}}.

\bibitem{Bustamante:2015waa}
M.~Bustamante, J.~F. Beacom, and W.~Winter, ``{\em {Theoretically palatable
  flavor combinations of astrophysical neutrinos}},''
  \href{http://dx.doi.org/10.1103/PhysRevLett.115.161302}{Phys. Rev. Lett.
  {\normalfont \bfseries 115} (2015)  161302},
\href{http://arxiv.org/abs/1506.02645}{{\normalfont \ttfamily
  arXiv:1506.02645}}.

\bibitem{Vincent:2016nut}
A.~C. Vincent, S.~Palomares-Ruiz, and O.~Mena, ``{\em {Analysis of the 4-year
  IceCube high-energy starting events}},''
  \href{http://dx.doi.org/10.1103/PhysRevD.94.023009}{Phys. Rev. {\normalfont
  \bfseries D94} (2016)  023009},
\href{http://arxiv.org/abs/1605.01556}{{\normalfont \ttfamily
  arXiv:1605.01556}}.

\bibitem{Poulin:2016nat}
V.~Poulin, P.~D. Serpico, and J.~Lesgourgues, ``{\em {A fresh look at linear
  cosmological constraints on a decaying dark matter component}},''
  \href{http://dx.doi.org/10.1088/1475-7516/2016/08/036}{JCAP {\normalfont
  \bfseries 1608} (2016)  036},
\href{http://arxiv.org/abs/1606.02073}{{\normalfont \ttfamily
  arXiv:1606.02073}}.

\bibitem{FrankiewiczonbehalfoftheSuper-KamiokandeCollaboration:2016pkv}
{\normalfont \bfseries Super-Kamiokande}, K.~Frankiewicz, ``{\em {Indirect
  searches for dark matter particles with the Super-Kamiokande detector}},''
\href{http://dx.doi.org/10.1393/ncc/i2015-15125-y}{Nuovo Cim. {\normalfont
  \bfseries C38} (2016) no.~4, 125}.

\bibitem{Dudas:2014bca}
E.~Dudas, Y.~Mambrini, and K.~A. Olive, ``{\em {Monochromatic neutrinos
  generated by dark matter and the seesaw mechanism}},''
  \href{http://dx.doi.org/10.1103/PhysRevD.91.075001}{Phys. Rev. {\normalfont
  \bfseries D91} (2015)  075001},
\href{http://arxiv.org/abs/1412.3459}{{\normalfont \ttfamily arXiv:1412.3459}}.

\bibitem{Abbasi:2011eq}
{\normalfont \bfseries IceCube}, R.~Abbasi {\em et al.}, ``{\em {Search for
  Dark Matter from the Galactic Halo with the IceCube Neutrino Observatory}},''
  \href{http://dx.doi.org/10.1103/PhysRevD.84.022004}{Phys. Rev. {\normalfont
  \bfseries D84} (2011)  022004},
\href{http://arxiv.org/abs/1101.3349}{{\normalfont \ttfamily arXiv:1101.3349}}.

\bibitem{Aisati:2015vma}
C.~El~Aisati, M.~Gustafsson, and T.~Hambye, ``{\em {New Search for
  Monochromatic Neutrinos from Dark Matter Decay}},''
  \href{http://dx.doi.org/10.1103/PhysRevD.92.123515}{Phys. Rev. {\normalfont
  \bfseries D92} (2015)  123515},
\href{http://arxiv.org/abs/1506.02657}{{\normalfont \ttfamily
  arXiv:1506.02657}}.

\bibitem{Ibarra:2013zia}
A.~Ibarra, A.~S. Lamperstorfer, and J.~Silk, ``{\em {Dark matter annihilations
  and decays after the AMS-02 positron measurements}},''
  \href{http://dx.doi.org/10.1103/PhysRevD.89.063539}{Phys. Rev. {\normalfont
  \bfseries D89} (2014)  063539},
\href{http://arxiv.org/abs/1309.2570}{{\normalfont \ttfamily arXiv:1309.2570}}.

\bibitem{Covi:2008jy}
L.~Covi, M.~Grefe, A.~Ibarra, and D.~Tran, ``{\em {Unstable Gravitino Dark
  Matter and Neutrino Flux}},''
  \href{http://dx.doi.org/10.1088/1475-7516/2009/01/029}{JCAP {\normalfont
  \bfseries 0901} (2009)  029},
\href{http://arxiv.org/abs/0809.5030}{{\normalfont \ttfamily arXiv:0809.5030}}.

\bibitem{Covi:2009xn}
L.~Covi, M.~Grefe, A.~Ibarra, and D.~Tran, ``{\em {Neutrino Signals from Dark
  Matter Decay}},'' \href{http://dx.doi.org/10.1088/1475-7516/2010/04/017}{JCAP
  {\normalfont \bfseries 1004} (2010)  017},
\href{http://arxiv.org/abs/0912.3521}{{\normalfont \ttfamily arXiv:0912.3521}}.

\bibitem{Fernandez:2015vhy}
{\normalfont \bfseries Super-Kamiokande}, P.~Fern{\'a}ndez, ``{\em {GADZOOKS!
  (SuperK-Gd): status and physics potential}},''
PoS {\normalfont \bfseries ICRC2015} (2016)  1131.

\bibitem{Lang:2016zhv}
R.~F. Lang, C.~McCabe, S.~Reichard, M.~Selvi, and I.~Tamborra, ``{\em
  {Supernova neutrino physics with xenon dark matter detectors: A timely
  perspective}},'' \href{http://dx.doi.org/10.1103/PhysRevD.94.103009}{Phys.
  Rev. {\normalfont \bfseries D94} (2016)  103009},
\href{http://arxiv.org/abs/1606.09243}{{\normalfont \ttfamily
  arXiv:1606.09243}}.

\bibitem{Audren:2014bca}
B.~Audren, J.~Lesgourgues, G.~Mangano, P.~D. Serpico, and T.~Tram, ``{\em
  {Strongest model-independent bound on the lifetime of Dark Matter}},''
  \href{http://dx.doi.org/10.1088/1475-7516/2014/12/028}{JCAP {\normalfont
  \bfseries 1412} (2014)  028},
\href{http://arxiv.org/abs/1407.2418}{{\normalfont \ttfamily arXiv:1407.2418}}.

\bibitem{DiValentino:2016foa}
{\normalfont \bfseries CORE}, E.~Di~Valentino {\em et al.}, ``{\em {Exploring
  Cosmic Origins with CORE: Cosmological Parameters}},''
\href{http://arxiv.org/abs/1612.00021}{{\normalfont \ttfamily
  arXiv:1612.00021}}.

\bibitem{Slatyer:2016qyl}
T.~R. Slatyer and C.-L. Wu, ``{\em {General Constraints on Dark Matter Decay
  from the Cosmic Microwave Background}},''
  \href{http://dx.doi.org/10.1103/PhysRevD.95.023010}{Phys. Rev. {\normalfont
  \bfseries D95} (2017)  023010},
\href{http://arxiv.org/abs/1610.06933}{{\normalfont \ttfamily
  arXiv:1610.06933}}.

\bibitem{Giesen:2015ufa}
G.~Giesen, M.~Boudaud, Y.~Génolini, V.~Poulin, M.~Cirelli, P.~Salati, and
  P.~D. Serpico, ``{\em {AMS-02 antiprotons, at last! Secondary astrophysical
  component and immediate implications for Dark Matter}},''
  \href{http://dx.doi.org/10.1088/1475-7516/2015/09/023,
  10.1088/1475-7516/2015/9/023}{JCAP {\normalfont \bfseries 1509} (2015)  023},
\href{http://arxiv.org/abs/1504.04276}{{\normalfont \ttfamily
  arXiv:1504.04276}}.

\bibitem{Boyarsky:2007ge}
A.~Boyarsky, D.~Malyshev, A.~Neronov, and O.~Ruchayskiy, ``{\em {Constraining
  DM properties with SPI}},''
  \href{http://dx.doi.org/10.1111/j.1365-2966.2008.13003.x}{Mon. Not. Roy.
  Astron. Soc. {\normalfont \bfseries 387} (2008)  1345},
\href{http://arxiv.org/abs/0710.4922}{{\normalfont \ttfamily arXiv:0710.4922}}.

\bibitem{Yuksel:2007dr}
H.~Yuksel and M.~D. Kistler, ``{\em {Circumscribing late dark matter decays
  model independently}},''
  \href{http://dx.doi.org/10.1103/PhysRevD.78.023502}{Phys. Rev. {\normalfont
  \bfseries D78} (2008)  023502},
\href{http://arxiv.org/abs/0711.2906}{{\normalfont \ttfamily arXiv:0711.2906}}.

\bibitem{Ackermann:2015lka}
{\normalfont \bfseries Fermi-LAT}, M.~Ackermann {\em et al.}, ``{\em {Updated
  search for spectral lines from Galactic dark matter interactions with pass 8
  data from the Fermi Large Area Telescope}},''
  \href{http://dx.doi.org/10.1103/PhysRevD.91.122002}{Phys. Rev. {\normalfont
  \bfseries D91} (2015) no.~12, 122002},
\href{http://arxiv.org/abs/1506.00013}{{\normalfont \ttfamily
  arXiv:1506.00013}}.

\bibitem{Cohen:2016uyg}
T.~Cohen, K.~Murase, N.~L. Rodd, B.~R. Safdi, and Y.~Soreq, ``{\em {Gamma-ray
  Constraints on Decaying Dark Matter and Implications for IceCube}},''
\href{http://arxiv.org/abs/1612.05638}{{\normalfont \ttfamily
  arXiv:1612.05638}}.

\bibitem{Ibarra:2013cra}
A.~Ibarra, D.~Tran, and C.~Weniger, ``{\em {Indirect Searches for Decaying Dark
  Matter}},'' \href{http://dx.doi.org/10.1142/S0217751X13300408}{Int. J. Mod.
  Phys. {\normalfont \bfseries A28} (2013)  1330040},
\href{http://arxiv.org/abs/1307.6434}{{\normalfont \ttfamily arXiv:1307.6434}}.

\bibitem{Essig:2013goa}
R.~Essig, E.~Kuflik, S.~D. McDermott, T.~Volansky, and K.~M. Zurek, ``{\em
  {Constraining Light Dark Matter with Diffuse X-Ray and Gamma-Ray
  Observations}},'' \href{http://dx.doi.org/10.1007/JHEP11(2013)193}{JHEP
  {\normalfont \bfseries 11} (2013)  193},
\href{http://arxiv.org/abs/1309.4091}{{\normalfont \ttfamily arXiv:1309.4091}}.

\bibitem{Cirelli:2012ut}
M.~Cirelli, E.~Moulin, P.~Panci, P.~D. Serpico, and A.~Viana, ``{\em {Gamma ray
  constraints on Decaying Dark Matter}},''
  \href{http://dx.doi.org/10.1103/PhysRevD.86.083506,
  10.1103/PhysRevD.86.109901}{Phys. Rev. {\normalfont \bfseries D86} (2012)
  083506},
\href{http://arxiv.org/abs/1205.5283}{{\normalfont \ttfamily arXiv:1205.5283}}.

\bibitem{Dugger:2010ys}
L.~Dugger, T.~E. Jeltema, and S.~Profumo, ``{\em {Constraints on Decaying Dark
  Matter from Fermi Observations of Nearby Galaxies and Clusters}},''
  \href{http://dx.doi.org/10.1088/1475-7516/2010/12/015}{JCAP {\normalfont
  \bfseries 1012} (2010)  015},
\href{http://arxiv.org/abs/1009.5988}{{\normalfont \ttfamily arXiv:1009.5988}}.

\bibitem{Cirelli:2009dv}
M.~Cirelli, P.~Panci, and P.~D. Serpico, ``{\em {Diffuse gamma ray constraints
  on annihilating or decaying Dark Matter after Fermi}},''
  \href{http://dx.doi.org/10.1016/j.nuclphysb.2010.07.010}{Nucl. Phys.
  {\normalfont \bfseries B840} (2010)  284--303},
\href{http://arxiv.org/abs/0912.0663}{{\normalfont \ttfamily arXiv:0912.0663}}.

\bibitem{Zhang:2009ut}
L.~Zhang, C.~Weniger, L.~Maccione, J.~Redondo, and G.~Sigl, ``{\em
  {Constraining Decaying Dark Matter with Fermi LAT Gamma-rays}},''
  \href{http://dx.doi.org/10.1088/1475-7516/2010/06/027}{JCAP {\normalfont
  \bfseries 1006} (2010)  027},
\href{http://arxiv.org/abs/0912.4504}{{\normalfont \ttfamily arXiv:0912.4504}}.

\bibitem{Aguilar:2016kjl}
{\normalfont \bfseries AMS}, M.~Aguilar {\em et al.}, ``{\em {Antiproton Flux,
  Antiproton-to-Proton Flux Ratio, and Properties of Elementary Particle Fluxes
  in Primary Cosmic Rays Measured with the Alpha Magnetic Spectrometer on the
  International Space Station}},''
\href{http://dx.doi.org/10.1103/PhysRevLett.117.091103}{Phys. Rev. Lett.
  {\normalfont \bfseries 117} (2016)  091103}.

\bibitem{Ade:2015xua}
{\normalfont \bfseries Planck}, P.~A.~R. Ade {\em et al.}, ``{\em {Planck 2015
  results. XIII. Cosmological parameters}},''
  \href{http://dx.doi.org/10.1051/0004-6361/201525830}{Astron. Astrophys.
  {\normalfont \bfseries 594} (2016)  A13},
\href{http://arxiv.org/abs/1502.01589}{{\normalfont \ttfamily
  arXiv:1502.01589}}.

\bibitem{Boddy:2015efa}
K.~K. Boddy and J.~Kumar, ``{\em {Indirect Detection of Dark Matter Using
  MeV-Range Gamma-Ray Telescopes}},''
  \href{http://dx.doi.org/10.1103/PhysRevD.92.023533}{Phys. Rev. {\normalfont
  \bfseries D92} (2015)  023533},
\href{http://arxiv.org/abs/1504.04024}{{\normalfont \ttfamily
  arXiv:1504.04024}}.

\bibitem{Knodlseder:2016pey}
J.~Kn{\"o}dlseder, ``{\em {The future of gamma-ray astronomy}},''
  \href{http://dx.doi.org/10.1016/j.crhy.2016.04.008}{Comptes Rendus Physique
  {\normalfont \bfseries 17} (2016)  663--678},
\href{http://arxiv.org/abs/1602.02728}{{\normalfont \ttfamily
  arXiv:1602.02728}}.

\bibitem{Feng:1997tn}
J.~L. Feng, T.~Moroi, H.~Murayama, and E.~Schnapka, ``{\em {Third generation
  familons, b factories, and neutrino cosmology}},''
  \href{http://dx.doi.org/10.1103/PhysRevD.57.5875}{Phys. Rev. {\normalfont
  \bfseries D57} (1998)  5875--5892},
\href{http://arxiv.org/abs/hep-ph/9709411}{{\normalfont \ttfamily
  arXiv:hep-ph/9709411}}.

\bibitem{Hirsch:2009ee}
M.~Hirsch, A.~Vicente, J.~Meyer, and W.~Porod, ``{\em {Majoron emission in muon
  and tau decays revisited}},''
  \href{http://dx.doi.org/10.1103/PhysRevD.79.055023}{Phys. Rev. {\normalfont
  \bfseries D79} (2009)  055023},
  \href{http://arxiv.org/abs/0902.0525}{{\normalfont \ttfamily
  arXiv:0902.0525}}.
[Erratum: Phys. Rev.D79,079901(2009)].

\bibitem{Jodidio:1986mz}
A.~Jodidio, B.~Balke, J.~Carr, G.~Gidal, K.~Shinsky, {\em et al.}, ``{\em
  {Search for Right-Handed Currents in Muon Decay}},''
\href{http://dx.doi.org/10.1103/PhysRevD.34.1967}{Phys. Rev. {\normalfont
  \bfseries D34} (1986)  1967}.

\bibitem{Albrecht:1995ht}
{\normalfont \bfseries ARGUS Collaboration}, H.~Albrecht {\em et al.}, ``{\em
  {A Search for lepton flavor violating decays $\tau\to e\alpha$,
  $\tau\to\mu\alpha$}},''
\href{http://dx.doi.org/10.1007/BF01579801}{Z. Phys. {\normalfont \bfseries
  C68} (1995)  25--28}.

\bibitem{Bayes:2014lxz}
{\normalfont \bfseries TWIST}, R.~Bayes {\em et al.}, ``{\em {Search for two
  body muon decay signals}},''
  \href{http://dx.doi.org/10.1103/PhysRevD.91.052020}{Phys. Rev. {\normalfont
  \bfseries D91} (2015)  052020},
\href{http://arxiv.org/abs/1409.0638}{{\normalfont \ttfamily arXiv:1409.0638}}.

\bibitem{Goldman:1987hy}
J.~T. Goldman, A.~Hallin, C.~Hoffman, L.~Piilonen, D.~Preston, {\em et al.},
  ``{\em {Light Boson Emission in the Decay of the $\mu^+$}},''
\href{http://dx.doi.org/10.1103/PhysRevD.36.1543}{Phys. Rev. {\normalfont
  \bfseries D36} (1987)  1543--1546}.

\bibitem{TheMEG:2016wtm}
{\normalfont \bfseries MEG}, A.~M. Baldini {\em et al.}, ``{\em {Search for the
  lepton flavour violating decay $\mu ^+ \rightarrow \mathrm {e}^+ \gamma $
  with the full dataset of the MEG experiment}},''
  \href{http://dx.doi.org/10.1140/epjc/s10052-016-4271-x}{Eur. Phys. J.
  {\normalfont \bfseries C76} (2016) no.~8, 434},
\href{http://arxiv.org/abs/1605.05081}{{\normalfont \ttfamily
  arXiv:1605.05081}}.

\bibitem{Cheng:1980tp}
T.~P. Cheng and L.-F. Li, ``{\em {$\mu \to e \gamma$ in Theories With Dirac and
  Majorana Neutrino Mass Terms}},''
\href{http://dx.doi.org/10.1103/PhysRevLett.45.1908}{Phys. Rev. Lett.
  {\normalfont \bfseries 45} (1980)  1908}.

\bibitem{Heurtier:2016iac}
L.~Heurtier and D.~Teresi, ``{\em {Dark matter and observable Lepton Flavour
  Violation}},'' \href{http://dx.doi.org/10.1103/PhysRevD.94.125022}{Phys. Rev.
  {\normalfont \bfseries D94} (2016)  125022},
\href{http://arxiv.org/abs/1607.01798}{{\normalfont \ttfamily
  arXiv:1607.01798}}.

\bibitem{Raffelt:1994ry}
G.~Raffelt and A.~Weiss, ``{\em {Red giant bound on the axion--electron
  coupling revisited}},''
  \href{http://dx.doi.org/10.1103/PhysRevD.51.1495}{Phys. Rev. {\normalfont
  \bfseries D51} (1995)  1495--1498},
\href{http://arxiv.org/abs/hep-ph/9410205}{{\normalfont \ttfamily
  arXiv:hep-ph/9410205}}.

\bibitem{Raffelt:2012sp}
G.~Raffelt, ``{\em {Limits on a CP-violating scalar axion--nucleon
  interaction}},'' \href{http://dx.doi.org/10.1103/PhysRevD.86.015001}{Phys.
  Rev. {\normalfont \bfseries D86} (2012)  015001},
\href{http://arxiv.org/abs/1205.1776}{{\normalfont \ttfamily arXiv:1205.1776}}.

\bibitem{Armengaud:2013rta}
E.~Armengaud {\em et al.}, ``{\em {Axion searches with the EDELWEISS-II
  experiment}},'' \href{http://dx.doi.org/10.1088/1475-7516/2013/11/067}{JCAP
  {\normalfont \bfseries 1311} (2013)  067},
\href{http://arxiv.org/abs/1307.1488}{{\normalfont \ttfamily arXiv:1307.1488}}.

\bibitem{Aprile:2014eoa}
{\normalfont \bfseries XENON100}, E.~Aprile {\em et al.}, ``{\em {First Axion
  Results from the XENON100 Experiment}},''
  \href{http://dx.doi.org/10.1103/PhysRevD.90.062009}{Phys. Rev. {\normalfont
  \bfseries D90} (2014)  062009},
\href{http://arxiv.org/abs/1404.1455}{{\normalfont \ttfamily arXiv:1404.1455}}.

\bibitem{Abe:2014zcd}
{\normalfont \bfseries XMASS}, K.~Abe {\em et al.}, ``{\em {Search for bosonic
  superweakly interacting massive dark matter particles with the XMASS-I
  detector}},'' \href{http://dx.doi.org/10.1103/PhysRevLett.113.121301}{Phys.
  Rev. Lett. {\normalfont \bfseries 113} (2014)  121301},
\href{http://arxiv.org/abs/1406.0502}{{\normalfont \ttfamily arXiv:1406.0502}}.

\bibitem{Abgrall:2016tnn}
{\normalfont \bfseries Majorana}, N.~Abgrall {\em et al.}, ``{\em {New limits
  on Bosonic Dark Matter, Solar Axions, Pauli Exclusion Principle Violation,
  and Electron Decay from the Majorana Demonstrator}},''
  \href{http://dx.doi.org/10.1103/PhysRevLett.118.161801}{Phys. Rev. Lett.
  {\normalfont \bfseries 118} (2017)  161801},
\href{http://arxiv.org/abs/1612.00886}{{\normalfont \ttfamily
  arXiv:1612.00886}}.

\bibitem{Rodejohann:2011mu}
W.~Rodejohann, ``{\em {Neutrino-less Double Beta Decay and Particle
  Physics}},'' \href{http://dx.doi.org/10.1142/S0218301311020186}{Int. J. Mod.
  Phys. {\normalfont \bfseries E20} (2011)  1833--1930},
\href{http://arxiv.org/abs/1106.1334}{{\normalfont \ttfamily arXiv:1106.1334}}.

\bibitem{Gando:2012zm}
{\normalfont \bfseries KamLAND-Zen}, A.~Gando {\em et al.}, ``{\em {Limit on
  Neutrinoless $\beta\beta$ Decay of $^{136}$Xe from the First Phase of
  KamLAND-Zen and Comparison with the Positive Claim in $^{76}$Ge}},''
  \href{http://dx.doi.org/10.1103/PhysRevLett.110.062502}{Phys. Rev. Lett.
  {\normalfont \bfseries 110} (2013)  062502},
\href{http://arxiv.org/abs/1211.3863}{{\normalfont \ttfamily arXiv:1211.3863}}.

\bibitem{Heurtier:2016otg}
L.~Heurtier and Y.~Zhang, ``{\em {Supernova Constraints on Massive
  (Pseudo)Scalar Coupling to Neutrinos}},''
  \href{http://dx.doi.org/10.1088/1475-7516/2017/02/042}{JCAP {\normalfont
  \bfseries 1702} (2017)  042},
\href{http://arxiv.org/abs/1609.05882}{{\normalfont \ttfamily
  arXiv:1609.05882}}.

\bibitem{Patel:2015tea}
H.~H. Patel, ``{\em {Package-X: A Mathematica package for the analytic
  calculation of one-loop integrals}},''
  \href{http://dx.doi.org/10.1016/j.cpc.2015.08.017}{Comput. Phys. Commun.
  {\normalfont \bfseries 197} (2015)  276--290},
\href{http://arxiv.org/abs/1503.01469}{{\normalfont \ttfamily
  arXiv:1503.01469}}.

\bibitem{Patel:2016fam}
H.~H. Patel, ``{\em {Package-X 2.0: A Mathematica package for the analytic
  calculation of one-loop integrals}},''
\href{http://arxiv.org/abs/1612.00009}{{\normalfont \ttfamily
  arXiv:1612.00009}}.

\bibitem{Binosi:2003yf}
D.~Binosi and L.~Theussl, ``{\em {JaxoDraw: A Graphical user interface for
  drawing Feynman diagrams}},''
  \href{http://dx.doi.org/10.1016/j.cpc.2004.05.001}{Comput. Phys. Commun.
  {\normalfont \bfseries 161} (2004)  76--86},
\href{http://arxiv.org/abs/hep-ph/0309015}{{\normalfont \ttfamily
  arXiv:hep-ph/0309015}}.

\end{thebibliography}\endgroup

\end{document}